\documentclass[AMA,STIX1COL]{WileyNJD-v2}
\usepackage[colorinlistoftodos]{todonotes}

\usepackage{siunitx}
\usepackage{chngcntr}
\usepackage{centernot}
\usepackage{xcolor}
\usepackage{pifont}
\usepackage{multirow}
\usepackage{lscape}
\usepackage{longtable}
\usepackage{setspace}

\newcommand\ind{\protect\mathpalette{\protect\independenT}{\perp}}
\def\independenT#1#2{\mathrel{\rlap{$#1#2$}\mkern2mu{#1#2}}}

\articletype{Tutorials in Biostatistcs}%
\received{<day> <Month>, <year>}
\revised{<day> <Month>, <year>}
\accepted{<day> <Month>, <year>}

\begin{document}

\title{Using principal stratification in analysis of clinical trials}

\author[1]{Ilya Lipkovich}
\author[2]{Bohdana Ratitch}
\author[1]{Yongming Qu}
\author[3]{Xiang Zhang}
\author[1]{Mingyang Shan}
\author[4]{Craig Mallinckrodt}

\authormark{LIPKOVICH \textsc{et al}}

\address[1]{ \orgname{Eli Lilly and Company}, \orgaddress{\city{Indianapolis}, \state{Indiana}, \country{USA}}}
\address[2]{ \orgname{Bayer}, \orgaddress{\city{Montreal}, \state{QC}, \country{Canada}}}
\address[3]{ \orgname{CSL Behring}, \orgaddress{\city{King of Prussia}, \state{PA}, \country{USA}}}
\address[4]{ \orgname{Pentara}, \orgaddress{\city{Millcreek}, \state{UT}, \country{USA}}}

\corres{Ilya Lipkovich,  Eli Lilly and Company, Indianapolis, IN 46285, USA\\
\email{ilya.lipkovich@lilly.com}}

\abstract[Abstract]{
The ICH E9(R1) addendum (2019) proposed principal stratification (PS) as one of five strategies for dealing with intercurrent events. Therefore, understanding the strengths, limitations, and assumptions of PS is important for the broad community of clinical trialists. Many approaches have been developed under the general framework of PS in different areas of research, including experimental and observational studies. These diverse applications have utilized a diverse set of tools and assumptions. Thus, need exists to present these approaches in a unifying manner. The goal of this tutorial is threefold. First, we provide a coherent and unifying description of PS. Second, we emphasize that estimation of effects within PS relies on strong assumptions and we thoroughly examine the consequences of these assumptions to understand in which situations certain assumptions are reasonable. Finally, we provide an overview of a variety of key methods for PS analysis and use a real clinical trial example to illustrate them. Examples of code for implementation of some of these approaches are given in supplemental materials.
}

\maketitle

\section{Introduction}\label{introduction}
How should we characterize the effect of a treatment on quality of life in a study where many patients die? How can we evaluate in a parallel design the efficacy of an experimental treatment vs control within the subset of patients who would tolerate the experimental drug? These are some of the many circumstances in which \textit{principal stratification} (PS) methodology can be useful.

Principal stratification partitions the clinical study population into latent sub-populations (principal strata). The partitioning is based on potential outcomes (PO) of a post-randomization variable (e.g., compliance to the assigned treatment) that lies on the causal pathway between the treatment and the outcome of primary interest. The objective of PS is to draw inference on treatment effects within principal strata, referred to as principal effects. Often, one of the principal strata is the focus of inference, but sometimes it is of interest to combine principal effects across several (or all) principal strata while accounting for a confounding effect of a post-randomization variable (e.g. see Egleston et al., 2010\cite{egleston2010tutorial}). 

The PS methodology emerged in 1980-90's (Robins, 1986\cite{robins1986}; Angrist and Imbens, 1995 \cite{angrist_imbens1995}; Angrist, Imbens and Rubin, 1996 \cite{air1996identification}). The clinical questions that motivated initial applications of PS centered on the effects of non-compliance and estimating treatment effects in the subgroup of patients that would be compliant with all treatments in the trial. The idea was to define causal estimands that would provide an alternative to the pure Intention-to-Treat (ITT) approach to estimate treatment effects in the relevant sub-population (stratum) of patients. 

Principal strata are defined in alignment with the sub-populations of interest, such as with respect to treatment compliance or changes in the originally assigned treatment regimen. Other applications of PS include stratification by various post-randomization outcomes/events rather than changes in treatment regimen (see review papers by VanderWeele, 2011\cite{vanderweele2011principal}, Mealli et al., 2012 \cite{Mealli2012}, and Bornkamp et al., 2021 \cite{bornkamp2021}). For example: 

\begin{itemize}
	\item Analysis of endpoints truncated by death, such as evaluating treatment effect on quality of life among survivors (patients who would survive regardless of assigned treatment)
	\item Treatment effect in patients who would experience no serious adverse events (AEs) on experimental treatment (regardless of their actual treatment received or if assigned to an experimental treatment)
	\item Treatment effect in a sub-population defined by a post-randomization event where the outcome of interest is measurable only in patients who had an event. For example, in vaccine studies, the effect of treatment on symptom severity can be evaluated only in those patients who would be infected regardless of randomized treatment
	\item Treatment effect in a sub-population defined by a post-randomization event that confounds the treatment effect, e.g., when the objective is to evaluate treatment effect within a patient subset defined by a risk factor or a surrogate of the outcome of interest
    \item Direct effect of treatment on an outcome variable in the sub-population of patients whose intermediate outcome(s) are unaffected by the treatment (``principal strata direct effect'' a.k.a ``dissociative effects''), similar to \textit{direct effect} in mediation analysis\cite{vanderweele2011principal, Mealli2012, Rubin2004, VanderWeele2009, Kim2019}. This application of PS is outside the scope of this tutorial.
\end{itemize}

The ICH E9(R1) addendum \cite{international2020harmonised} introduced the term \textit{intercurrent event} (ICE), which is defined as a post-randomization event that ``affect either interpretation or existence of the measurements associated with clinical questions of interest''. Examples of ICEs include post-randomization events such as switching to a different treatment, treatment discontinuation/incompliance, or death. Such events inherently complicate the interpretation of outcomes. For example, for a patient who discontinued assigned treatment, interpretation of a causal link between their outcomes and the assigned treatments may be complicated even if this patient still remained in the trial and provided outcomes at scheduled clinical visits. The addendum suggested five strategies for dealing with intercurrent events for causal estimands. Principal stratification is one of the strategies mentioned in ICH E9(R1), which fostered a wave of interest from  clinical trialists in the theory and application of PS-based estimands and estimators. We stress, that while forming strata based on ICEs is one popular application of PS, our review considers PS based on any post-baseline  variable and is not limited to ICEs (e.g, it can be based on levels of a surrogate biomarker).

This tutorial serves several purposes. First we unify diverse literature on applications of PS scattered across different research communities. Often, methods are motivated by a narrow problem and it may not be immediately clear that an approach can be generalized for other tasks. As a result, very similar approaches may be developed in different contexts, using different language and notation, thereby leading to confusion and misunderstanding. Second, we discuss common estimation issues in the PS framework, emphasizing that the estimation of principal effects always requires assumptions that cannot be verified from observed data. We examine commonly used assumptions and their analytic implications. We further consider situations in which the various assumptions may or may not be plausible. Finally, we provide an overview of a variety of key methods for PS analysis and their implementation using an example based on a real clinical trial data. We make the implementation of these analyses publicly available by providing either the corresponding code or references to the developer's resources.

The tutorial is organized as follows. A formal framework for principal stratification, as introduced in Frangakis and Rubin (2002) \cite{frangakis2002principal} relies on the notion of potential outcomes (POs), which we review in Section~\ref{background}. Section~\ref{data.example} introduces a clinical trial example that is used throughout the paper. Section~\ref{ps} introduces two common settings for PS, based on adherence and post-randomization outcomes. Section~\ref{common.assumptions} considers some common assumptions in the PS methods that are considered in later sections. Section~\ref{estimators.3assum} describes the one historically significant CACE (Complier Average Causal Effect) estimator. Section~\ref{estimators.2assum} reviews several methods for estimating PS effects based on the monotonicity assumption. Section~\ref{sc.ps.monotonicity} reviews  sensitivity analysis frameworks that relax the monotonicity assumption. Section~\ref{sc.ps.SACE} reviews methods for estimating bounds on PS effects. Section~\ref{sc.partial.compliance} reviews literature on estimating principal causal effects (PCE) for strata based on the joint distribution of partial compliances. Section~\ref{sc.ps.baseline} introduces approaches that utilize baseline covariates. Section~~\ref{sc.ps.baseline.post-baseline} reviews recently proposed methods that utilize baseline and post-baseline covariates using direct likelihood based estimation and multiple imputation. Section~\ref{sc.ps.bayesian} introduces methods based on Bayesian joint modeling of principal strata and potential outcomes. Section~\ref{sc.ps.nonrand} reviews ideas for extending PS estimation to non-randomized trials. Finally, Section~\ref{sc.summary} concludes with a summary and discussion. 

\section{Background -- Notation and Potential Outcomes}\label{background}
In this section we introduce notation and the key concept of \textit{potential outcomes} that are used throughout this tutorial.

We assume a randomized (controlled) clinical trial (RCT) with two treatment arms, with $T_i$ denoting the indicator for the treatment arm, with $T_i=0$ for control and $T_i=1$ for experimental treatment, where $i=1,2,\ldots,n$ is the index across subjects (we use ``subject'' and ``patient'' interchangeably). The outcome of interest (endpoint) for subject $i$, denoted as $Y_i$, can be a continuous or binary variable. We limit our considerations to binary and continuous outcomes to avoid complexities with time to event analysis that may obscure general ideas. Baseline and post-baseline covariates are denoted as $X_i$ and $Z_i$, respectively (as random variables for \textit{i}th subject). These can be matrices with row vectors $\textbf{X}_i=(X_{i1},\ldots, X_{ip})$ and $\textbf{Z}_i=(Z_{i1},\dots ,Z_{ik})$. While in the examples we use multiple covariates in $X$ and $Z$, in the notation, whenever possible, we refer to a single covariate to avoid boldface type. 

In general, capital letters denote random variables, whether observable or potential outcomes, or columns of data matrices. Realized values of observable variables and variables designating (dummy) arguments of functions are denoted with small letters. For example, $Y_i$ and $Y_i(t)$ denote the observable and potential outcomes for the $i$th patient (introduced in the following paragraph) as random variables when used in theoretical considerations such as evaluating expected value of an estimator. Because such random variables are exchangeable across patients, often the patient index will be dropped unless confusion may arise. Conversely, $y_i$ denotes the observed (realized) value for $i$th subject in the data set that is used in describing computational formulas of estimators. 
 
The concepts presented in this tutorial rely on the concept of potential outcomes. The idea behind POs comes from Neyman (1923) \cite{neyman1923} and was reinforced in Rubin (1974)\cite{Rubin1974}; that is, for every patient in a clinical trial, there is an outcome on each candidate treatment that could potentially be observed. Of course, in a parallel group trial each patient can be observed on only one treatment. Similar to the outcome of interest $Y_i$, POs can be defined with respect to any  post-randomization variable, outcome, event, or change in treatment. In general the potential outcome under treatment $T=t, t \in \{0,1\}$is denoted by placing the treatment as an argument, $(t)$, following the variable name. For example, $Y_i(0)$ denotes the potential outcome for the response variable $Y$ for subject $i$ assuming this patient would have received control treatment ($T=0$). PO's are often (somewhat inaccurately) referred to as \textit{counterfactuals} because they apply to outcomes for actually assigned treatments (factuals) and to the hypothetical treatments that the patients could have received, but was not assigned to. That is, the potential outcome $Y(1-t)$ on treatment $1-t$ counter to the fact that the patient was randomized to treatment $t$. Therefore, strictly speaking, there are two potential outcomes in a dichotomous treatment assignment scenario, one factual and one counterfactual.

Some other important examples of post-randomization events $S \in \{0,1\}$ are: a direct \textit{outcome} of treatment, such as an AE, lack of efficacy, infection, or an outcome derived from continuous biomarker(s) exceeding pre-specified cutoff(s); or a change in the \textit{treatment regimen} that may be related to early outcomes of the treatment, whether planned (e.g., initiation of concomitant medication, rescue, or switch to an alternative treatment), or spontaneous (e.g., lack of compliance with the assigned treatment and study protocol). We use $S$ generically for post-baseline variables defining principal strata. For some methods of PS, we use other post-baseline variables $Z$ that are predictive of strata membership. Often $Z$s are earlier measures of the primary outcome $Y$.

The fundamental assumption that allows us to connect potential and observed outcomes at an individual patient level is the \textit{consistency assumption}, implied by a more general \textit{stable unit treatment value assumption} (SUTVA) (see Rubin, 1980 \cite{rubin1980}): 
\begin{equation}
\begin{aligned}
& \nonumber \mbox{A1: } \; Y_i=Y_i (1) T_i+Y_i (0)(1-T_i) \\
& \nonumber \mbox{A2: } \; S_i=S_i (1) T_i+S_i (0)(1-T_i) \\
& \nonumber \mbox{A3: } \; Z_i=Z_i (1) T_i+Z_i (0)(1-T_i).
\end{aligned}
\end{equation}

In words, the observed outcome is the same as (consistent with) the potential outcome associated with the treatment that the subject was assigned/randomized to. 

An important consequence of a randomized assignment to treatment groups $T \in \{0,1\}$ is that POs are independent of the randomized treatment assignment $T$. For example, the distribution of potential outcomes for a patient hypothetically assigned to active treatment, $Y_i(1)$, does not depend on the treatment $T_i$ that the patient was actually randomized to. More generally, a consequence of randomized treatment assignment is independence of the assigned treatment from the joint distribution of all PO's. 
Symbolically,
\[
\mbox{A4: } \; T_i \ind \{Y_i(0),Y_i(1),S_i(0),S_i(1),Z_i(0),Z_i(1)\}.
\]

Although the concept of PO's has been used predominantly in causal literature as a tool for fostering inference from observational data, PO's are a natural language for defining estimands in general, thus serving equally well randomized and observational studies. Note that estimands are typically defined as expectations of individual-level contrasts. For example, the average treatment effect (ATE) estimand is the  expectation over $Y_i(1)-Y_i(0)$ for a binary or continuous $Y$. In RCTs, the ATE causal estimand can be expressed via expectations of observable outcomes (assuming compliance to initially randomized treatment)
\begin{equation}\label{eq.ATE}
\begin{aligned}
 E(Y(1)-Y(0)) & =  E(Y(1))-E(Y(0)) \\
 &= E(Y(1)|T=1)-E(Y(0)|T=0)\\
 &= E(Y|T=1)-E(Y|T=0).
\end{aligned}
\end{equation}
The second line follows from the treatment ignorability assumption A4 and the third line\textemdash  from the consistency assumption A1. It then follows that the ATE can be consistently estimated in RCTs by the difference in the mean estimates between 2 treatment groups as long as we can consistently estimate expected outcomes within each treatment arm: $E(Y|T=t), t \in {0,1}$.

Although the population relevant for most estimands is the full study population (all randomized patients), sometimes, we need to consider an estimand defined for a sub-population. Principal stratification provides a way of defining sub-populations of interest based on post-randomization outcomes/events without compromising causality and dealing with the potentially confounding nature of post-randomization variables.

Frangakis and Rubin (2002) \cite{frangakis2002principal} developed a formal framework for PS where patient memberships in principal strata is defined in terms of their potential outcomes $S_i(0)$ and $S_i(1)$ (see Section \ref{ps} for a detailed discussion). The authors argued that estimates of treatment effect within patient subsets formed by principal strata are causal because potential outcomes $S_i(t)$, determining strata membership, are independent of treatment assignment. Therefore, $S_i(t)$ can be considered similar to values of a pre-treatment covariates upon which treatment effect can be conditioned. This can be contrasted with various attempts on estimands formed by conditioning on actual post-randomization outcomes within each arm. Such estimands do not maintain  randomization and are not causal. A notoriously popular example is ``completer's'' (often called ``per-protocol'') analysis that compares summaries across treatments conditioned on completing the trial without protocol deviations. 

The difficulty with conditioning on potential outcomes, however, is that $S_i(t)$ are typically only partially observed in a parallel arm RCT: for each patient, the only potential outcome that can be observed is the one associated with the randomized treatment actually assigned to the subject. Therefore, additional identifying assumptions are needed for estimating treatment effects within principal strata. Indeed, as we will see, a major challenge of using principal stratification for analysis of data from clinical trials is its counterfactual nature that requires strong assumptions to be able to identify and estimate treatment effect within principal strata.

We emphasize that all PS methods rely on strong and untestable assumptions (for RCTs with parallel treatment groups) and therefore require sensitivity analyses against departures from these assumptions. Many PS methods come with explicit sensitivity parameters that represent such untestable assumptions, so each  assumption can be linked with a specific value or range of values of the parameter. As a result, estimates of stratum-specific effects can be reported as an interval of estimates under plausible range of underlying sensitivity parameters. However, some of the methods do not have explicit sensitivity parameters instead relying on various (untestable) ``ignorability'' assumptions that take a form of independence of certain potential outcomes conditional on observed data or on other potential outcomes, which may be hard to conceptualize. Sensitivity analyses against such assumptions are not straightforward typically requiring introducing unmeasured confounders causing violation of the above assumptions.

In causal literature\cite{Robins1998, daniel2015}, potential outcomes are sometimes associated with joint application of (intervention on) the initial randomized treatment $T$ and post-randomization outcomes/treatment changes $S$. To avoid additional notation, we use $S$ to denote any post-randomization event, which may not necessarily be a stratification variable). Adopting their notation, $Y_i(t,s)$ is the potential outcome for an arbitrary subject $i$ who was randomized to treatment $T=t$ but actually treated (or later switched to) treatment/condition $S=s$. This notation assumes that it is possible to intervene on both $T$ and $S$ (i.e., set specific conditions $T=t$ and $S=s$). In situations where $S$ means an outcome, such as an adverse event rather than a change in treatment regimen, $Y_i(t,s)$ with $s \neq S_i(t)$ may be unrealistically counterfactual (such potential outcomes are sometimes referred to as \textit{a priori} counterfactuals\cite{Mealli2012, Kim2019}). For example, you cannot induce an AE in a subject by directly intervening with the adverse event. Importantly, within the PS framework we do not need to assume such ``extremely counterfactual'' outcomes, because the goal is to consider potential outcomes conditioned on specific and relevant levels of $\{S(0),S(1)\}$. For example, when evaluating the expected value $E(Y(t,s)\mid S(t)=s)$ we can assume that, by the \textit{composition} assumption\cite{VanderWeele2009}, for the subset of subjects with $S(t)=s$, we should observe $Y(t,s)=Y(t,S(t)) \equiv Y(t)$. In words, we consider an intervention on $T$ only, by setting $T=t$, while $S$ is set at the same value $S(t)$ that would have been observed for that patient had s/he been randomized to $T=t$, and, therefore, the effect of intervening on $S$ can be ignored; that is, $E(Y(t,s)\mid S(t)=s)=E(Y(t)\mid S(t)=s)$. For example, assume we look at patients for whom under the active treatment $(T=1)$ there is no AE $(S(1)=0)$, so $E(Y(1,s)|S(1)=s) = E(Y(1,0)|S(1)=0)\equiv E(Y(1)|S(1)=0)$. Since we condition on a hypothetical subset of patients who would not have AE if treated, their observed AE status will be $S=0$ and by the composition assumption we can remove the second argument from $Y(1,0)$. Therefore, to simplify, only single-dimensional indexing of POs is used throughout this tutorial.

The presented notation is essential and sufficient for defining many principal stratification estimands that have been (and continue to be) considered in the literature. However, there are important extensions that go beyond this framework. One limitation of our  notation is that it does not reflect the longitudinal aspect of clinical trials. This may be relevant for PS in several ways. First, when the outcome is time to event, estimating effects within principal strata may require joint modeling of time to the event defining the primary outcome and the event defining PS. Our review is limited to binary and continuous outcomes. Secondly, for longitudinal studies with continuous or binary outcomes, estimating effects within PS may require joint modeling of outcome and stratification variables as repeated measures. Most PS methods were developed under the ``fixed-time'' strata and considerable work may be required to extend them to longitudinal settings. Our review follows the original development of each method and therefore adopts a ``fixed-time'' setting in most places except Section~\ref{sc.ps.baseline.post-baseline}, which presents approaches that explicitly examine repeated measures for estimating PS effects. Thirdly, in addition to repeated outcomes, PS framework can incorporate time-varying  treatments and compliance\cite{Frangakis2004, Lin2008, Dai2012}. In particular, strata can be defined at different stages of a multi-stage trial with patients being re-randomized at the beginning of each stage with different treatment options available depending on earlier outcomes (sequential, multiple assignment, randomized trials, SMART). Stratification variables (e.g., based on stage-wise compliance) can be defined for SMART trials targeted at estimating treatment effects associated with dynamic treatment regimens (DTR) for patients within PS based on such variables. For example, we may be interested in PS defined by a subset of patients who would be compliant to a certain combination of assigned treatments at the first and second stages. Here we do not consider PS effects associated with DTR and refer an interested reader to recent research\cite{Artman2020, Bhattacharya2021} and references therein.  

Another limitation of this review is that we focus on the settings where PS is based on binary potential outcomes $S(t) \in \{0,1\}, t\in \{0,1\}$ defined prior to estimation of treatment effects. In contrast, there is a stream of literature\cite{Kim2019,jin2008principal,Bartolucci2011} that aims at simultaneously estimating treatment effects within all possible subsets formed by conditioning on any specific values of continuous potential outcomes $D(0),D(1)$. As a typical example that motivated such extensions, $D$ can be \textit{partial compliance} measured as the proportion of the assigned dose (e.g., proportion of the total number of pills) of an experimental drug or control taken during the clinical trial, with interest on estimating treatment effect within strata defined by levels of $D(0),D(1)$. This topic is discussed briefly in Section~\ref{sc.partial.compliance}.


\section{Diabetes example}\label{data.example}
Throughout this article we will use a data set based on the IMAGINE-3 Study: a 52-week, multi-center, phase 3 study of patients with type 1 diabetes mellitus. This was a parallel, double-blind study with randomization of qualified subjects to basal insulin lispro (BIL) versus insulin glargine (GL) \textemdash two long-acting insulin formulations, with addition of short-acting insulin used for controlling the postprandial glucose level or for correcting high glucose at any time. In this trial, 1114 adults were randomized to BIL or GL in a 3:2 ratio (664 in BIL: 450 in GL), stratified by baseline hemoglobin A1c (HbA1c) ($\le$8.5\%, >8.5\%), baseline low-density lipoprotein cholesterol (LDL-C) (<100 mg/dL [2.6 mmol/L], $\ge$100 mg/dL), and prior basal insulin therapy (GL/insulin detemir/other basal insulin). Insulin doses were adjusted weekly in the first 12 weeks of treatment and then adjusted according to investigators' judgement thereafter. Patients were not allowed to take additional anti-diabetes medication unless they discontinued the randomized study treatment. 

The primary objective of the clinical trial was to demonstrate the non-inferiority of BIL to GL on HbA1c after 52 weeks of treatment and the superiority was tested in a sequential manner if the non-inferiority was met. This study was registered at \texttt{https://clinicaltrials.gov} as NCT01454284 and details of the study have been published \cite{bergenstal2016randomized}. Recently, this data set was used to illustrate the principal stratification approach for evaluating treatment effect in principal strata based on compliance when postbaseline intermediate outcomes are considered \cite{qu2020general, qu2020imp}. Here we will also use compliance as the post-randomization variable of interest to define PS.  

\section{Defining principal strata}\label{ps}

\subsection{Defining principal strata based on post-randomization treatment adherence}\label{sc.ps.trtchange}
The PS strategy can be useful for estimands in sub-populations defined by changes in treatment regimen that typically occur in response to early treatment outcomes. For example, consider compliance/ non-compliance with the initially assigned (randomized) treatment. Some may argue that compliance is an outcome of treatment rather than part of a pre-specified treatment regime. However, compliance/non-compliance is a decision\textemdash a choice of treatment made by patients/doctors/caregivers. These decisions can be contrasted with treatment outcomes such as adverse events or worsening symptoms that are events experienced by patients and not decisions about treatments (interventions) that are \textit{applied} to patients.

In the literature on RCTs, ``adherence''  is mostly used as a binary indicator for whether a patient was taking the assigned treatment whereas ``compliance'' is mostly used to quantify how many doses the patient missed while still continuing with treatment. However, in this article we use the two terms interchangeably, as the causal literature often refers to ``compliance'' in the same context as clinical trialists refer to ''adherence''.

Consider the situation in which the control treatment is placebo. Arguably lack of compliance with the experimental treatment effectively switches  non-compliant patients to placebo, or at least we might reasonably expect ``placebo-like'' outcomes. Although this assumption of equivalence of ``no treatment'' to ``placebo'' motivated early research on causal estimands for RCTs with noncompliance\cite{littlerubun2000causal}, this may be an obvious oversimplification that trivializes the complex role of placebo arms, including controlling for the ``placebo effect'' in RCTs.  Symmetrically, lack of compliance to control (placebo) treatment may mean taking the alternative (experimental) treatment. This setting may have been suggested by post-marketing trials and observational studies where patients have access to alternative treatments. As an example, the REFLUX trial \cite{grant2013REFLUX} comparing a laparoscopic surgery with a noninvasive treatment allowed physician to switch treatment for certain patients from that assigned by randomization to the alternative treatment based on clinical considerations. Also some may consider ``non-compliers'' those patients who were rescued to an active treatment due to poor outcomes on placebo. However, this can be questioned arguing that such patients are ``compliers'' if they precisely followed a pre-defined regimen that includes rules for rescuing. Nevertheless this setting is historically significant as it motivated the first applications of PS where the objective was to estimate the Complier Average Causal Effect (CACE), also known as Local Average Treatment Effect (LATE); see, for example, Angrist, Imbens, and Rubin (1996) \cite{air1996identification}, Imbens and Rubin (1997) \cite{imbens1997bayesian}.

In the context considered in this section, we denote the actual (or adopted) treatment of $i$th patient by $S_i$, where (consistent with coding for the assigned treatment $T_i$) $S_i=1$ means experimental treatment and $S_i=0$ control (in the context of an RCT). For each patient we define two potential outcomes $S_i(t) \in \{0,1\}, t=0,1$ that gives rise to various principal strata. For example, $S_i(1)=0$ represents patients who, if randomized to experimental treatment, would be taking control at the end of the trial when the outcome of primary interest $Y$ is measured. Similarly, $S(0)=0$ represents patients who, if randomized to control, would continue taking control through the end of the trial. Based on this coding we can label the four principal strata as illustrated in Table~\ref{tab:compl} where rows represent POs on a control treatment, $S(0)$, and columns\textemdash on an experimental treatment, $S(1)$. Here, the compliant stratum of patients is formed by those in the set $\{i: S_i(0)=0,S_i(1)=1\}$. Subject subscripts can usually be removed from the notation of potential and observed outcomes without causing confusion, especially when the potential and observed outcomes are treated as random variables from a sample of exchangeable units, e.g., observable outcomes $Y,S$ or potential outcomes $Y(0),Y(1),S(0),S(1)$.Using this notation, the CACE estimand defines a treatment effect within the principal stratum of \textit{Compliers}, 
\[
\text{CACE}=E[Y(1)-Y(0) \mid S(1)=1,S(0)=0].
\]

In Section~\ref{sc.ps.monotonicity} we will consider a more general notion of compliance that, in our opinion, is more applicable for the analysis of RCT. There, adherence/compliance to the assigned treatment simply means that patients take the assigned medication throughout the study. Lack of adherence/compliance therefore does not imply switching to an alternative treatment (i.e. that a patient who was randomized to a control treatment takes an experimental treatment). Therefore, in Section~\ref{sc.ps.monotonicity} we use different notation where the intercurrent event $S=1$ indicates lack of compliance (non-compliance) and $S=0$ indicates compliance. In this case, the principal stratum of patients who would adhere to both treatments is designated as $\{S(0)=0, S(1)=0\}$, that is subjects who would comply with both the experimental and control treatments if assigned to those treatments. 

\begin{table}[b] \centering
\caption{Principal strata defined by post-randomization treatment $S$ ($S=1$ if taking experimental treatment; $S=0$ if taking control).\label{tab:compl}}
\begin{tabular}{lcc}
\hline
& $S(1)=0$ & $S(1)=1$\\
\addlinespace[0.2cm]
\hline
$S(0)=0$ & Never-takers &  Compliers \\
$S(0)=1$ & Defiers & Always-takers \\
\hline  
\end{tabular}
\end{table}

It is important to note that the \textit{Compliers} principal stratum is not the same as a subset of subjects in a clinical trial who are observed to be compliant with the one treatment to which they were randomized. For example, a subject randomized to control and observed to be compliant may or may not be compliant with the experimental treatment has s/he been randomized to it. Consequently, comparison of outcomes $Y$ between subjects observed to comply with the control treatment versus subjects observed to comply with the experimental treatment is not a causal estimand because it compares two different populations that may differ regarding important prognostic characteristics. In contrast, comparing treatments within the \textit{Compliers} principal stratum represents a causal estimand because the treatments are compared within the same patient sub-population, where the membership in this stratum is defined based on potential outcomes, $S(0)=0$ and $S(1)=1$, which are independent of treatment assignment. The CACE estimand can also be of interest, for example, in equivalence studies. Here CACE would be a more principled approach than a Per Protocol analysis, in which inclusion in the analysis is determined from observed compliance status on each patient's randomized treatment\cite{lou2019estimation, lou2019ratio}.

When comparing an active treatment to placebo, interest may be in evaluating the effect in patients who would be able to comply with an active treatment, regardless of compliance to placebo; such as, when compliance to placebo is irrelevant in real clinical settings. In this case, focus is on the treatment effect within the subset $S(1)=1$ combining two principal strata\textemdash \textit{Compliers} and \textit{Always-takers}. In other situations, interest may be in the subset of patients who would not be able to tolerate placebo regardless of their compliance with the active treatment. Because these patients cannot tolerate ``no treatment'' (placebo) they are potentially in greatest need of active intervention. In such cases, the focus is on the treatment effect within the set $S(0)=1$ combining  \textit{Defiers} and \textit{Always-takers}. Principal stratification based on PO framework is flexible and can be applied to diverse settings. As an example, in Section~\ref{sc.ps.baseline.post-baseline} we provide approaches for estimating treatment effect in the stratum of patients who can tolerate the experimental treatment regardless of treatment assignment, $S(1)=0$.

\subsection{Defining principal strata based on post-randomization outcomes}\label{sc.ps.outcomes}

Now we consider a situation where stratification is based on a post-randomization event $S$ that does not represent treatment (compliance/adherence) status but rather an event such as adverse reaction, death, worsening of disease severity or other early outcome associated with treatment (hereafter referred to generically as "event''). In this case, we want to estimate treatment effect in a stratum defined by occurrence (or non-occurrence) of the event under study treatments, i.e., $S(0)$ and $S(1)$. A special case of post-randomization event is when the outcome of interest can be measured only in patients who experience a specific post-randomization event, e.g., those who were infected, had a certain surgical procedure, etc.

The principal strata defined in this context are illustrated in Table~\ref{tab:symptom}, which is similar to Table~\ref{tab:compl}, except now our post-randomization event represents a presence or absence of a harmful outcome such as relapse, hence, the four cells are labeled differently. Depending on the nature of the event $S$, different principal strata may be of interest. For example, if the event is worsening of a disease-related outcome, we may want to estimate the treatment effect in a stratum of patients who are immune to this event (i.e. would not experience relapse) no matter what treatment they are randomized to (i.e, those with $S(1)=S(0)=0$, where $S=1$ is indicating relapse). This may be of interest, for example, in long-term treatment decisions if disease worsening can be anticipated based on patient characteristics or determined relatively early after treatment initiation. 

\begin{table}[b]\centering
\caption{Principal strata defied by a post-randomization event $S$, with $S=1$ indicating a negative treatment outcome, such as relapse.\label{tab:symptom}}
\begin{tabular}{lcc}
\hline
 & $S(1)=0$ & $S(1)=1$\\
\addlinespace[0.2cm]
\hline
$S(0)=0$ & Immune (I) &  Harmed (H) \\
$S(0)=1$ & Benefiters (B) & Doomed (D) \\
\hline  
\end{tabular}
\end{table}

If experiencing the event is required for measuring the primary outcome $Y$, the stratum of interest may be those who would experience the event, no matter what treatment they are randomized to  (e.g, those with $S(1)=S(0)=1$, where  $S=1$ is indicating a post-randomization event). For example, in a vaccine study where the viral load is the primary outcome and is only measurable in HIV-infected patients (who were not-infected at randomization and became infected post-randomization).

Another example is the Survivor Average Causal Effect (SACE) that has been discussed in Zhang and Rubin (2003) \cite{zhang2003estimation} in the context of estimating causal treatment effects on a clinical or quality of life outcomes when the outcome may be truncated by death. Let $S(t)=0$ denote whether a patient survives to a time point of interest under treatment $t=0,1$, there are four principal strata as described in Table~\ref{tab:survival}. Interest is in the treatment effect in the principal stratum of patients who would survive to a time point of interest on either treatment (\textit{Always-survivors}):  
\[
\text{SACE}=E[Y(1)-Y(0) \mid S(1)=S(0)=0],
\]
where $S(t)=1$ represents death (the intercurrent event) and $S(t)=0$ survival to a pre-defined time point when the outcome $Y$ is measured under treatment $t=0,1$. 

\begin{table}[t] \centering
\caption{Principal strata defined by post-randomization survival status ($S=0$ indicating survival).\label{tab:survival}}
\begin{tabular}{lcc}
\hline
 & $S(1)=0$ & $S(1)=1$\\
\addlinespace[0.2cm]
\hline
$S(0)=0$ & Always-survivors &  Control-only survivors\\
$S(0)=1$ & Experimental-only survivors & Doomed\\
\hline  
\end{tabular}
\end{table}

\section{Common assumptions used for estimation}\label{common.assumptions}
As mentioned in Section~\ref{background}, the estimation of treatment effects within principal strata requires knowledge of strata membership for all study subjects, which is only partially observable in parallel-arm studies. To deal with this partial observability when estimating principal effects, certain assumptions need to be made, as is always the case in presence of missing data. Most assumptions aim at eliminating some unobserved components or at postulating equality between potential outcomes in some strata so that the unobserved potential outcomes can be estimated by observed outcomes. In this section, we summarize several common assumptions and in the following sections we  discuss estimation approaches that rely on these assumptions.

Let the true proportions of patients in the four cells of Table~\ref{tab:compl} corresponding to \textit{Never-takers}, \textit{Compliers}, \textit{Defiers}, and \textit{Always-takers}, be $\pi_{00},\pi_{01},\pi_{10},\pi_{11}$. Although these probabilities are unknown, the marginal probabilities are estimable from the observed data. For example, $\textrm{Pr(}S(0)=0)=\pi_{00} + \pi_{01}$ can be estimated in an unbiased manner as the proportion of patients randomized to placebo who remained on placebo at the end of the study. Ability to estimate marginal probabilities combined with assumptions such as outlined below allows estimation of principal effects. 

In addition to the SUTVA assumption (needed to connect PO's with observables) discussed in Section \ref{background}, the following assumptions are sometimes made:

\begin{enumerate}
 \item	\textit{Exclusion restriction}: assumes that potential outcomes for \textit{Never-takers} and \textit{Always-takers} are the same, regardless of what arm they were randomized to, i.e., $Y(1) = Y(0)$ in these two principal strata where $\{S(1)=S(0)\}$;
\item	\textit{Monotonicity}: $S(1) \ge S(0)$, which means that there are no \textit{Defiers}, i.e., the stratum $\{S(0)=1, S(1)=0 \}$ is empty and, therefore, $\pi_{10}=0$.
\item	\textit{Positivity}: the probability of membership in the stratum of interest is positive, (e.g., if the focus is on \textit{Compliers}, $\pi_{01}>0$).
\end{enumerate}

The exclusion restriction may be more easily justifiable for the setting of Table~\ref{tab:compl} as patients for whom $S(0)=S(1)$ effectively means taking the same  treatment, hence the same potential outcomes $Y(0)=Y(1)$. In other contexts of PS, e.g., when evaluating treatment effect within strata defined by symptom worsening/improvement or an adverse event, the exclusion restriction assumptions are less natural, essentially assuming ``no treatment effect'' for a stratification variable implies ``no treatment effect'' in the primary outcome. In the example of Table~\ref{tab:symptom}, this  means the equality of potential outcomes $(Y(1)=Y(0))$ in both \textit{Immune} and \textit{Doomed} strata. In other words, the entire or full treatment effect must be mediated via the intermediate (stratification) variable. In other cases, such as considered in Table~\ref{tab:survival}, and the case of $S$ being a generic indicator for compliance to the initial treatment (see Table~\ref{tab:compl2} of Section~\ref{sc.ps.monotonicity}), this may be an overly strong assumption. 

The implications of the monotonicity assumption in the example when the ICE is lack of compliance are that compliance with the experimental treatment is assumed to be at least as good as or better than compliance with the control treatment. The direction of the monotonicity assumption depends on whether the stratification outcome is favorable or not. When principal strata are defined based on an unfavorable post-randomization outcome (e.g., death), this assumption means that if a patient has  this outcome on the experimental treatment then s/he would also experience this outcome on the control treatment. In other words, it is assumed that the experimental treatment cannot harm subjects in terms of the stratification outcome $S$ more than the control treatment can. This condition implies that the stratum of \textit{Control-only} survivors (in the terminology of Table~\ref{tab:survival}) is empty and can be algebraically expressed as  $S(1) \le S(0)$. On the other hand, if the stratification outcome is favorable, such as remission, the monotonicity relation is reversed as $S(1) \ge S(0)$.

Therefore, monotonicity is a strong assumption, especially its deterministic version and it often may be replaced with stochastic monotonicity\cite{small2017stochastic}. Another route for not relying on  monotonicity is via incorporating data on baseline and post-baseline covariates, which  also requires additional assumptions (see Section~\ref{sc.ps.baseline.post-baseline}). A simple and often taken approach for relaxing monotonicity is at the expense of introducing additional sensitivity parameter(s) as described in detail in Section~\ref{sc.ps.monotonicity}.

Returning to the example in Table~\ref{tab:compl} where lack of compliance means switching to the other treatment, we note that in some trials patients who are non-compliant with placebo would have no access to an active drug. When this is the case, a stronger assumption can be made that the \textit{Always-takers} stratum has a zero probability of occurrence. Sometimes it is referred to as ``one-sided non-compliance'' (see, e.g., Ding and Lu, 2017\cite{ding2017principal}).  

Several estimation methods rely on modeling to identify strata membership based on observed data, e.g., baseline covariates. The key assumption in this case is that of \textit{principal ignorability} (PI; see Jo and Stuart, 2009 \cite{jo2009use}) which states that the observed covariates are sufficient for identifying principal stratum membership. A strong version of the PI assumption can be expressed in terms of independence of the potential outcomes and strata membership given covariates, implying that expected potential outcomes are the same in all strata, given covariates:
\begin{equation}\label{eq.PI}
\begin{aligned}
Y(1) \ind (S(0),S(1))\mid X, \\
Y(0) \ind (S(0),S(1))\mid X.
\end{aligned}
\end{equation}

This is similar to using covariates for propensity-based methods with observational data (see  Rosenbaum and Rubin, 1983 \cite{rosenbaum1983propscore};  Lunceford and Davidian, 2004 \cite{lunceford2004ps}) where potential outcomes are assumed to be independent of non-randomly assigned treatment $T$ given covariates:
\begin{equation}
\begin{aligned}
\nonumber
Y(1) \ind T \mid X, \\
Y(0) \ind T \mid X.
\end{aligned}
\end{equation}

Propensity score methods are discussed in detail in Section~\ref{sc.ps.nonrand}. Note that unlike the \textit{treatment ignorability} (TI) assumption, PI postulates independence between ``cross-world'' potential outcomes that inhabit "parallel universes" (such as, $Y(t)$ and $S(1-t)$) rather than between counterfactuals and observables ($Y(t)$ and $T$). Weaker versions of the PI assumption have been used by some methods in combination with other assumptions\cite{ding2017principal, feller2017principal}, as illustrated by the following equations: 
\begin{eqnarray}
\nonumber
Y(1) \ind S(0)\mid X, S(1)=1, \\
\nonumber
Y(0) \ind S(1)\mid X, S(0)=0.
\end{eqnarray}

In words, under the Weak PI, given covariates: ($i$) the distribution of potential outcomes under the experimental treatment, $Y(1)$, is the same in \textit{Always-takers} and \textit{Compliers} (i.e. for $S(1)=1$), and ($ii$) the distribution of potential outcomes under control, $Y(0)$, is the same in \textit{Never-takers} and \textit{Compliers}, (i.e. for $S(0)=0$).


Another assumption that can be useful in combination with PI is cross-world conditional independence of the stratification status $S(0)$ and $S(1)$ given baseline covariates, see Hayden et al. (2005)\cite{hayden2005estimator} who linked these assumptions with ``explainable nonrandom noncompliance'' of Robins (1988) \cite{Robins1998}.
\begin{eqnarray}\label{eq.crossworld.ind}
S(1) \ind S(0)\mid X.
\end{eqnarray}

The assumption presented in Equation (\ref{eq.crossworld.ind}) is particularly strong, essentially assuming that the cross-world random effects associated with the same patient are conditionally independent given baseline covariates, which like any other cross-world assumptions cannot be verified from the data when each patient receives only one treatment.

We already noted similarly between various assumptions of treatment ignorability conditional on covariates in causal inference of observational data and ignorability of strata conditional on covariates in PS analysis of randomized trials. The analogy provides insight to various PS methods employing covariates described in Sections \ref{sc.ps.baseline} and \ref{sc.ps.baseline.post-baseline}. In the analysis of observational data, TI lends to various standardization (more generally, g-estimation) strategies\cite{hernan2020book} that employ conditioning PO's on confounders (making it legitimate to replace PO's with observable outcomes under conditional expectation), followed by averaging conditional effects over the distribution of covariates in the overall population. Likewise, estimating effects in principal strata often requires similar ignorability assumptions.

The following sections provide an overview of the key types of methods for estimating treatment effects in the principal stratification framework.

\section{Estimators using the exclusion restriction, monotonicity, and positivity assumptions}\label{estimators.3assum}
Consider an estimand where the treatment effect of interest is the expected difference between  outcomes $Y(1)$ and $Y(0)$ in the principal stratum of Compliers (C) as defined in Table \ref{tab:compl}: 
\[
E(Y(1)-Y(0)|S(0)=0,S(1)=1) = E(Y(1)-Y(0)|C).
\]
Estimation can be accomplished using the exclusion restriction, monotonicity, and positivity assumptions. Per the monotonicity assumption, there are no \textit{Defiers} (D) and per the exclusion restriction, the treatment difference in \textit{Never-takers} (N) and \textit{Always-takers} (A) is 0.  Therefore, we can write the overall treatment effect as a weighted average of expected outcomes across the four strata (C, N, A, D):
\begin{eqnarray}
\nonumber E(Y(1)-Y(0)) &=& E(Y(1)-Y(0)\mid N) \times \pi_{00}  +  E(Y(1)-Y(0) \mid C)\times \pi_{01}\\ &+& 
\nonumber E(Y(1)-Y(0)\mid D) \times \pi_{10} + E(Y(1)-Y(0)\mid A) \times \pi_{11}  \\ &=&
\nonumber 0 \times \pi_{00} + E(Y(1)-Y(0)\mid C) \times \pi_{01} + E(Y(1)-Y(0)\mid D) \times 0 + 0 \times \pi_{11}. 
\end{eqnarray}

From the above and the positivity assumption, we can estimate the treatment effect in the \textit{Compliers} stratum using the overall treatment effect estimated from observed outcomes: 

\begin{equation}\label{eq.IV}
E(Y(1)-Y(0) \mid C)=\frac{E(Y(1))-E(Y(0))}{\pi_{01}}. 
\end{equation}

The numerator of (\ref{eq.IV}) can be estimated via a common estimator using observed outcomes based on the randomization and SUTVA assumptions as shown in (\ref{eq.ATE}). The denominator is identifiable by subtraction using the monotonicity assumption ($\pi_{10}=0$) and marginal probabilities which are estimable from observed data. Specifically, denote $p_0$ and $p_1$ the proportions of patients in the control and experimental arm, respectively, who would comply with their randomized treatment. It is easy to see that $p_1$ is a consistent estimator of the probability of \textit{Compliers} or \textit{Always-takers} (the second column of Table~\ref{tab:compl}), whereas assuming no \textit{Defiers} under monotonicity, $1-p_0$ is a consistent estimator of the probability of \textit{Always-takers} (the second row  of Table~\ref{tab:compl}). Therefore, we can consistently estimate the  probability of \textit{Compliers}, $\widehat{\pi}_{01} = p_1+p_0-1$. Expression (\ref{eq.IV}) is sometimes called the instrumental variable (IV) estimator of CACE \cite{air1996identification, littlerubun2000causal, Baiocchi2014}. 

\section{Estimators using monotonicity and positivity assumptions}\label{estimators.2assum}

In more general situations, e.g., when defining principal strata based on post-randomization outcomes (see Section \ref{sc.ps.baseline.post-baseline}), we often cannot rely on the ``exclusion restriction'' that greatly simplifies estimation by setting to zero the treatment effect in two principal strata. As a result, the remaining assumptions would not be sufficient to estimate all the necessary components, and some additional "sensitivity" parameters must be introduced, which we describe in this section. 

For this discussion consider the scenario represented by Table \ref{tab:symptom} and assume that  $S=1$ means poor efficacy, such as an early worsening of the condition under treatment. The estimand of interest is the treatment effect in the \textit{Immune} (I) stratum (the upper left cell). In this context, the monotonicity assumption is formulated as $S(1) \le S(0)$. This means that everyone who had an early disease worsening on active treatment would have also had early worsening  if randomized to placebo. Therefore, there are by assumption no \textit{Harmed} (H) patients ($S(0)=0,S(1)=1$) and this stratum can be considered empty, $\pi_{01}=0$.

To make a connection with the case where the stratification outcome $S$ is an adverse event, the monotonicity assumption would be reversed, $S(1) \ge S(0)$, implying that everyone who would have an AE on placebo would have also had it on active treatment, making the \textit{Benefiters} (B) an empty stratum. The positivity assumption assumes a positive probability in the stratum of interest, \textit{Immune},  $\pi_{00}>0$. First, we consider a (simpler) case where the outcome variable is binary and then consider a continuous outcome. 

\subsection{The case of a binary outcome}\label{sc.ps.outcomes.binary}

Now consider estimating the treatment effect as an odds ratio in the \textit{Immune} stratum:
\[
OR(I)=\frac{\textrm{Pr(}Y(1)=1 \mid I)(1-\textrm{Pr(}Y(0)=1 \mid I))}{(1-\textrm{Pr(}Y(1)=1 \mid I))\textrm{Pr(}Y(0)=1 \mid I)}.
\]
The null hypothesis of interest is $OR(I) \ge 1$ against the alternative $OR(I) <1$ (assuming the desirable outcome is $Y=0$).

Because the \textit{Harmed} stratum is assumed empty we can easily estimate the probability of membership in the \textit{Immune} stratum from the marginal probability as well as the conditional probability of potential outcomes $Y(0)$, given stratum $I$ using observed data. 

To estimate strata membership, we can write
\[
\textrm{Pr(}I)=\textrm{Pr(}S(0)=0)=\textrm{Pr(}S(0)=0 \mid T=0)= \textrm{Pr(}S=0 \mid T=0).  
\]
The first equality is based on the monotonicity assumption. The second equality follows from the independence of potential outcomes and treatment assignment (by randomization). The third uses SUTVA and allows us to replace potential outcomes with observables. Finally, the last term can be estimated as a simple proportion within the control arm.

To estimate the probability $Y(0)=1$ (or more broadly, the expected value of potential outcome $Y(0)$) within the \textit{Immune} stratum we can write 
\[
E(Y(0) \mid I)=E(Y(0) \mid S(0) =0)=E(Y(0) \mid S(0)=0, T=0)=E(Y \mid S=0, T=0).
\]
Again, the first equation follows from the monotonicity assumption, the second \textemdash  from independence, and the last one from SUTVA. These are the building blocks that allow us to better understand the role of different assumptions and facilitate more complex derivations.

Now the challenge is in estimating the expected value of $Y(1)$ within the \textit{Immune} stratum because, without making additional assumptions, it is only possible to estimate the probability of $Y(1)$ across the combination of \textit{Immune} and \textit{Benefiters} strata, $\textrm{Pr(}Y(1)=1 \mid I \;or\; B)$. 

One strategy is to express $\textrm{Pr(}Y(1)=1\mid I)$ via a sensitivity parameter for the expected value of $Y(0)$ in the \textit{Benefiters} stratum: 
\[
\gamma=\textrm{Pr(}Y(1)=1 \mid B).
\]
and use it in the sensitivity analysis of the $OR(I)$. 

Because $\textrm{Pr(}Y(1)=1 \mid I \; or\;  B)=\textrm{Pr(}Y(1)=1 \mid I)\textrm{Pr(}I \mid I \; or \; B) + \textrm{Pr(}Y(1)=1 \mid B)\textrm{Pr(}B \mid I \; or\;  B)$,
we can express the probability of binary outcomes in the \textit{Immune} stratum, $\textrm{Pr(}Y(1)=1 \mid I)$, as follows (e.g., see Magnusson et al., 2019 \cite{magnusson2019bayesian})
\[
\textrm{Pr(}Y(1)=1 \mid I)=\frac{\textrm{Pr(}Y(1)=1 \mid I \; or\;  B)}{\textrm{Pr(}I \mid I \; or \;  B)} - \frac{\textrm{Pr(}B \mid I\; or \;B)}{\textrm{Pr(}I \mid I \;or\; B)} \times \gamma,
\]

Note that all quantities except the sensitivity parameter $\gamma$ can be estimated from the observed data, therefore we can write the estimated $\widehat{OR}_\gamma(I)$ as a function of $\gamma$ and do stress-testing by varying $\gamma$ within a plausible range, assuming we can set a meaningful range other than the full interval from 0 to 1.

One can choose different sensitivity parameters\cite{egleston2010tutorial,egleston2007}, for example, the ``risk ratio'' for the probability of outcome $Y=1$ if treated in the \textit{Benefiters} stratum vs. \textit{Immune} stratum:
\[
\tau=\frac{\textrm{Pr(}Y(1)=1 \mid B)}{\textrm{Pr(}Y(1)=1 \mid I)}.
\]
The quantity of interest can then be expressed as 
\begin{equation}\label{eq.sens.tau}
\textrm{Pr(}Y(1)=1 \mid I)=\frac{\textrm{Pr(}Y(1)=1 \mid I \; or\; B)}{\tau+(1-\tau)\textrm{Pr(}I \mid I \; or\; B)}.
\end{equation}
An alternative specification of the sensitivity parameter, suggested by Gilbert, Bosch and Hudgens (2003), \cite{gbh2003sensitivity} (see also Mehrotra, Li, Gilbert, 2006 \cite{mlg2006comparison}), is as follows. We can write
\[
\textrm{Pr(}Y(1)=1 \mid I)=\frac{\textrm{Pr(}Y(1)=1\mid I \; or \; B)\textrm{Pr(}I \mid Y(1)=1,I \; or \; B)}{\textrm{Pr(}I \mid I \; or \; B)}.
\]
The only unidentifiable quantity is $P(I \mid Y(1)=1,I \;or \; B)$ which can be expressed via a sensitivity parameter $\beta$ using a logistic specification 
\begin{equation}\label{GBH.binary}
\textrm{Pr(}I \mid Y(1)=y,I \; or \; B)= \frac{1}{1+\exp(-\alpha - \beta y)},
\end{equation}
where $y \in \{0,1\}$.

When $\beta$ is fixed at any specific value, the parameter $\alpha$ can be identified from the constraint 
\[
\frac{\textrm{Pr(}I \; or\;  B)}{\textrm{Pr(}I)}\sum_{y=0}^{1}{\textrm{Pr(}I\mid Y(1)=y,I \; or \; B)\textrm{Pr(}Y(1)=y \mid I \; or \; B)} =1.
\]
One can argue that specifying a sensitivity parameter $\beta$ is more convenient than working with $\gamma$ because it has the following natural interpretation as a log odds ratio. Essentially, it captures our understanding of how well the outcome on drug for an event-free patient predicts what would happen if that same patient were randomized to control. For example, if $\beta <0$, then those event-free patients in the treated arm ($S(1)=0$) with the better outcome $(Y=0)$ are more likely than those with the worse outcome $(Y=1)$ to be also event-free if randomized to placebo (i.e. being \textit{Immune}). 
\subsection{The case of a continuous outcome}\label{sc.ps.outcomes.continuous}

Now consider principal strata as in Table~\ref{tab:symptom}, except that the outcome variable $Y$ is continuous with the lower values meaning better outcomes. We follow the framework of Gilbert, Bosch and Hudgens (2003) \cite{gbh2003sensitivity} and Mehrotra et al. (2006) \cite{mlg2006comparison}. As before, interest is in evaluating treatment effect within the \textit{Immune} stratum, $I=\{S(0)=0, S(1)=0 \}$, which by monotonicity assumption is equivalent to $\{S(0)=0\}$. In the most general form, the null hypothesis can be formulated in terms of comparing distributions of potential outcomes within the \textit{Immune} stratum:
\[
F_0(y \mid I) = \textrm{Pr(}Y(0) \le y \mid I)\; vs. \; F_1(y \mid I)=\textrm{Pr(}Y(1) \le y \mid I), \forall  y.
\]
Similar to Section~\ref{sc.ps.outcomes.binary}, $F_0(y \mid I)$ can be estimated from the observed data in the placebo arm for patients with no event, $T=0,S=0$. However, $F_1 (y \mid I)$ cannot be estimated from observed data and requires a sensitivity parameter which (as in the binary case) can be expressed through the conditional distribution of $S(0)$ given $Y(1)$. Specifically, using a simple equation for conditional probably $\textrm{Pr(}A | B, C)=\textrm{Pr(}A | B)\frac{\textrm{Pr(}C | A, B)}{\textrm{Pr(}C | B)},$ and setting $A \equiv \{Y(1) = y\}$, $B \equiv \{S(1)=0\}$, $C \equiv \{S(0)=0\}$ we can write (under monotonicity):
\begin{eqnarray}
 f_{Y(1)}(y \mid S(1)=0,S(0)=0)  &=& 
f_{Y(1)}(y \mid S(1)=0)\frac{\textrm{Pr(}S(0)=0 \mid S(1)=0,Y(1)=y)}{\textrm{Pr(}S(0)=0 \mid S(1)=0)} 
\nonumber \\ &=& \frac{\textrm{Pr(}S(1)=0)}{\textrm{Pr(}S(0)=0)} f_{Y(1)}(y \mid S(1)=0)w(y),
\nonumber
\end{eqnarray}
where the weight $w(\cdot)$ is a function of potential outcome $Y(1)$, that can assume a logistic form 
\begin{equation}\label{eq.senspar1}
w(y)=\textrm{Pr(}S(0)=0 \mid S(1)=0, Y(1)=y)= \frac{1}{1+\exp(-\alpha - \beta y)}.
\end{equation}
The conditional density $f_{Y(1)}(y \mid S(1)=0)$ can be estimated from the observed data (parametrically or non-parametrically) by using observed $Y$ for patients randomized to the treatment arm $(T=1)$ who had no event $(S=0)$. Similarly, $\textrm{Pr(}S(1)=0)$ and $\textrm{Pr(}S(0)=0)$ are estimated as the proportions of patients without the symptom in the treated and control arm, respectively. The intercept $\alpha$ in the logistic model is identifiable for any $\beta$ from the constraint that the full integral for $f_1(y \mid I)$ is unity,
\begin{equation}\label{eq.alpha}
\frac{\textrm{Pr(}S(1)=0)}{\textrm{Pr(}S(0)=0)}\int_{-\infty}^{\infty}{f_{Y(1)} (y \mid S(1)=0)w(y)dy} =1.
\end{equation}
The slope parameter, $\beta$ is a sensitivity parameter that quantifies the log odds for not having the event ($S=0$) if (hypothetically) randomized to the placebo arm, given someone was event-free and randomized to the treatment arm, per one unit change in outcome. Ideally, a plausible range for this parameter would be elicited from subject matter experts \cite{shepherd2007elicitation}.

The following statistic for testing the equality of distributions $F_0(y \mid I)$  and $F_1(y \mid I)$ based on the above formulation is adopted from Gilbert, Bosch, and Hudgens (2003) \cite{gbh2003sensitivity} (see also Lu, Mehrotra and Shepherd (2013) \cite{lms2013rankbased}: 
\begin{equation}\label{eq.gen.stat}
T(\beta)= \int_{-\infty}^{\infty}y\left(d\widehat{F}_0(y \mid I)- d\widehat{F}_1(y;\beta \mid I)\right).
\end{equation}
In a fully non-parametric setting the integral is a summation over observed patients: 
\begin{equation}\label{eq.GBH}
T(\beta) =\frac{1}{n_0} \sum_{i=1}^{N}{(1-t_i)(1-s_i)y_i} - \frac{\hat{p}_1}{\hat{p}_0} \frac{1}{n_1} \sum_{i=1}^{N}{t_i(1-s_i)w(y_i; \hat{\alpha}, \beta)y_i},
\end{equation}
where $N$ is the total number of subjects in the study, $n_0, n_1$ are the number of patients with $S=0$ in placebo and treated groups, respectively, and $\hat{p}_0$ and $\hat{p}_1$ are the proportions of patients without the event $S$ in the control and treated arms, respectively. 

For any specified value of $\beta, T(\beta)$ can be considered a bias-corrected version of a test statistic comparing group means in patients who had no event, where treated patients who have higher probability of no event if randomized to placebo are up-weighted, and patients with smaller probability are down-weighted. The standard errors can be computed using bootstrap. Lu, Mehrotra and Shepherd (2013) \cite{lms2013rankbased} also proposed a rank-based version of $T(\beta)$.

\subsubsection{Applying sensitivity analysis for completers to the diabetes example}\label{sc.ps.outcomes.continuous.example}

Here we apply the sensitivity analysis of Section~\ref{sc.ps.outcomes.continuous} to the data example from Section~\ref{data.example}. The outcome ($Y$) is defined as the difference from baseline to a 52-week endpoint in hemoglobin A1c (HbA1c), with larger negative values indicating improvement. The principal stratum of interest is based on completers (adherers) $S=0$, and we aim to estimate an average treatment effect in the ``Always-compliers'' (here ``Always-completers'') stratum, that is, $\{S(1) = 0, S(0)=0\}$. The observed proportion of patients who completed assigned treatment were 509/663 =76.7\% for BIL (the experimental drug, $T=1$) and 368/449=82.0\% for GL (the control, $T=0$). Therefore, the control patients were more likely to complete their assigned treatment. A naïve comparison of mean outcomes for patients who completed their treatments favored the experimental treatment ($-0.504$ for BIL vs. $-0.238$ for GL). We note that in this data set few patients ($n=11$) who completed the study without intercurrent events had missing HbA1c. For simplicity, our analysis is based on observed data, assuming missingness completely at random.

For PS methods requiring monotonicity, we assume that completing treatment on BIL implies being able to complete GL. In other words, the principal strata of those who would complete BIL but discontinue on GL is assumed to be empty. Under this assumption we apply a method developed in Gilbert, Boschand and Hudgens (2003) \cite{gbh2003sensitivity}, often referred to as GBH, implemented in R package \textbf{sensitivityPStrat}. The setting of this example is slightly different from that of Section~\ref{sc.ps.outcomes.continuous} as the monotonicity constraint is  now $S(1) \ge S(0)$, therefore the stratum $\{S(1)=0, S(0)=0\} \equiv \{S(1)=0\}$. Given that, the weight function from Eq.~(\ref{eq.senspar1}) will change as 
\[
w(y)=\textrm{Pr(}S(1)=0 \mid S(0)=0, Y(0)=y)= \frac{1}{1+\exp(-\alpha - \beta y)}.
\]

In words, $\beta$ expresses the log odds ratio of completing the experimental treatment per unit change in outcome $Y$ for patients who were actually assigned to and completed control treatment, if they were hypothetically assigned to the experimental treatment. We vary $\beta$ within a non-positive range from $-1.5$ to $0$ (in terms of odds ratios, from 0.223 to 1) assuming the larger negative values of $Y$ (indicating clinical benefit) increase the odds of completion on the alternative arm. For example, a unit drop of A1c assuming  $\beta = -1.5$ would result in $1/0.223=4.48$-fold increase in odds for completion on the alternative arm. Note that $\alpha$ is estimated for every assumed value of $\beta$ as explained in (\ref{eq.alpha}). 

\begin{figure}[ht]
  \centering
  \includegraphics[scale=0.63]{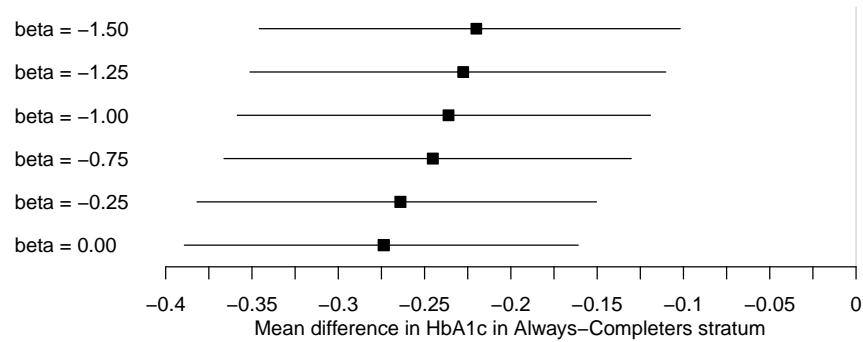}
  \caption{ Sensitivity analysis for diabetes data using the GBH method.}
  \label{fig.ex1.GBH}
\end{figure}

The point estimates for treatment effect within the PS stratum are obtained as a function of $\beta$  from a test statistic similar to (\ref{eq.GBH}) with an obvious change that the weight function is now applied to the control rather than treated patients. These and the associated 95\% confidence intervals (shown as horizontal lines) are plotted in Fig~\ref{fig.ex1.GBH}. The lower and upper confidence limits were obtained as 2.5 and 97.5 percentiles of test statistics evaluated on 1000 bootstrap samples. Computations were done using the R function \textit{sensitivityGBH}.

As the negative sensitivity parameter $\beta<0$ becomes larger by absolute value, the treatment effect shrinks. Numerically, this is because under larger negative values of $\beta$, the control patients with better outcomes receive larger relative weights in the test statistics. Conceptually, this follows from the basic idea of the sensitivity parameter: to increase chances for patients who completed the control treatment with good outcomes \textit{to also be completers} if assigned to the experimental arm. This is equivalent to selecting control patients with better than average outcomes to be compared with completers on the experimental arm, thus making it harder for the experimental arm to ``win''.  

\section{Relaxing the monotonicity assumption}\label{sc.ps.monotonicity}

Monotonicity may be plausible when active drug is compared with placebo. However, even in this case, strict (or deterministic monotonicity) may be implausible (see Small and Tan, 2017 \cite{small2017stochastic}; Qu et al., 2020 \cite{qu2020general}; Qu et al., 2020 \cite{qu2021mon}). For example, a patient may have no adverse reaction, e.g., weight gain, when exposed to treatment that tends to cause weight gain, yet gain weight due to other reasons when off treatment. Clearly, monotonicity is entirely implausible in bio-equivalence studies when the two treatments are almost identical (under alternative hypothesis). In general, relaxing the monotonicity assumption  requires introducing additional sensitivity parameters. Examples include Shepherd et al. (2011) \cite{shepherd2011sensitivity} and an approach considered in this section.

A simple and general sensitivity framework that does not assume monotonicity was proposed in Chiba and VanderWeele (2011) \cite{chiba2011simple} in the context of evaluating the average effect in ``survivors'' and was later used in Lou, Jones and Wanjie (2019) \cite{lou2019estimation} in bio-equivalence studies to assess the average effect in compliers. Here we follow \cite{lou2019estimation} to illustrate the evaluation of treatment effect in the stratum of patients who are compliant with either of the two active treatments. Therefore, we are considering a setting similar to that of Section~\ref{estimators.3assum}, but now we do not assume monotonicity.  

Here, in contrast with Table~\ref{tab:compl}, we used notation with $S=1$ for the intercurrent event of non-compliance to the initially assigned treatment and $S=0$ for compliance. This is a common notation in PS methods for compliance  arising in applications of RCT with an active comparator where the setting of Section \ref{estimators.3assum} may be too restrictive. For example, now we do not assume that non-compilers switch treatment in a parallel arm study, for example, non-compliance to assigned treatment may simply means taking no treatment. This setting is represented in Table~\ref{tab:compl2}.

\begin{table}[b]\centering
\caption{Principal strata defined by post-randomization treatment $S$ ($S=0$, if compliant with assigned treatment; $S=1$, if not).\label{tab:compl2}}
\begin{tabular}{lcc}
\hline  
 & $S(1)=0$ & $S(1)=1$\\
\addlinespace[0.2cm]
\hline  
$S(0)=0$ & Always-compliers &  Control-only-compliers\\
$S(0)=1$ & Experimental-only-compliers& Never-compilers\\
\hline  
\end{tabular}
\end{table}

The idea is to represent the effect in \textit{Compliers} as the ``observed effect in patients who complied'' (i.e., completed study on their randomized arm) $+  \alpha$, where $\alpha$ is a bias (sensitivity) parameter or a function of several sensitivity parameters that reflect different aspects of bias in the treatment effect in \textit{Compliers} that can be expected when an ``observed case'' analysis is used. Formally, 
\begin{equation}\label{eq.sens.CACE}
\begin{aligned}
 \delta &= E[Y(1) \mid S(0)=0,S(1)=0]-E[Y(0) \mid S(0)=0,S(1)=0] \\ &=  
 E[Y(1) \mid S(1)=0]- E[Y(0) \mid S(0)=0]+ \alpha.
 \end{aligned}
\end{equation}

The sensitivity parameter $\alpha$ expressing the bias of the "observed completers analysis" for the CACE estimand is actually a function of three sensitivity parameters, $\pi_{10}, \beta_0, \beta_1$.
\[
 \alpha=\frac{\pi_{01}}{\pi_0}\beta_0 - \frac{\pi_1-\pi_0+\pi_{01}}{\pi_1} \beta_1,
\]
where $\pi_0= \textrm{Pr(}S(0)=0)$, $\pi_1= \textrm{Pr(}S(1)=0)$ are marginal probabilities that can be estimated from data under randomization and the consistency assumption; 
$\pi_{01}=\textrm{Pr(}S(0)=0, S(1)=1)$, probability of  membership in the \textit{Control-only-compliers} stratum (recall $\pi_{01}=0$ under monotonicity, see Section~\ref{sc.ps.trtchange} where \textit{Defiers} was the zero probability stratum).

The slope parameters express the expected shift in potential outcomes under treatment (control) in those who comply only with experimental (control) treatment versus the outcomes in those who comply with both treatments:
\begin{eqnarray}
\nonumber \beta_1=E[Y(1) \mid S(1)=0,S(0)=1]- E[Y(1) \mid S(1)=0,S(0)=0] \\
\nonumber \beta_0=E[Y(0) \mid S(1)=1,S(0)=0]- E[Y(0) \mid S(1)=0,S(0)=0].
\end{eqnarray}
In the next section, we will use the ideas of quantifying bias with sensitivity parameters in the context of evaluating SACE and setting the lower and upper bounds on within-strata casual effects.  

\section{Estimating bounds on causal effects}\label{sc.ps.SACE}

A crude estimate of SACE has been proposed in Chiba and VanderWeele (2011) \cite{chiba2011simple}, which is expressed as the treatment effect estimated from the observed survivors (indicated with $S=0$) minus a sensitivity parameter (see also Section~\ref{sc.ps.monotonicity}) : 
\[
SACE=E[Y(1) \mid T=1, S=0] - E[Y(0) \mid T=0, S=0] - \alpha,
\]
where $\alpha =E[Y(1) \mid T=1,S=0]-E[Y(1) \mid T=0,S=0]$. The sensitivity parameter $\alpha$ represents the average difference in the outcome that would have been observed under the experimental treatment, $Y(1)$, comparing two populations: the first is the population that would have survived on the experimental treatment $(T=1,S=0)$, the second is the population that would have survived without the experimental treatment $(T=0,S=0)$. This is a conservative estimate of SACE under the following assumptions:
\begin{itemize}
    \item 	Monotonicity: $S(1) \le S(0)$ for all patients, i.e., survival under the experimental treatment is at least as good as under the control treatment and there is no heterogeneity of the treatment effect on survival. This assumption renders the cell of \textit{Control-only-Survivors} empty in Table~\ref{tab:survival}.
    \item 	$\alpha \le 0$, i.e., that the subset of survivors under the control treatment would have better outcomes on the experimental treatment than the population of survivors under the experimental treatment. In other words, it assumes that the control treatment survivors are healthier overall than the experimental treatment survivors and that the experimental treatment would never worsen their outcomes $Y(1)$.
\end{itemize}

The sensitivity parameter $\alpha$ would ideally be specified based on expert opinion and is not estimated from data. 

The approach described in Zhang and Rubin (2003) \cite{zhang2003estimation} and further justified by Kosuke (2008) \cite{imai2008sharp} (see a recent application in Colantuoni et al.\cite{colantuoni2018}) provides lower and upper bounds on a crude estimate of SACE under certain assumptions:
\[
Lower_{SACE} \le E[Y(1) \mid S(1)=S(0)=0] - E[Y(0) \mid S(1)=S(0)=0] \le Upper_{SACE},
\]
The average outcome in the control group, $E[Y(0) \mid S(1)=S(0)=0]$, can be estimated from the observed survivors under the control treatment based on the monotonicity assumption, i.e., that the control treatment survivors are also experimental treatment survivors ($S(1) \le S(0)$). The average outcome in the experimental group, $E[Y(1) \mid S(1)=S(0)=0]$ can be bounded as follows. The observed experimental survivors are a mix of \textit{Always-Survivors} and \textit{Experimental-only-Survivors}. The sharp bounds can be derived by assuming that \textit{Always-Survivors} would have better outcomes than those who would die under treatment or control (called ``ranked average score'' assumption in Zhang and Rubin, 2003 \cite{zhang2003estimation} and ``stochastic dominance'' in Kosuke, 2008 \cite{imai2008sharp}). Adding the monotonicity assumption allows for even sharper bounds. Under both assumptions, the lower bound can be estimated from observed experimental survivors, $E[Y(1) \mid T=1,S(1)=0]=E[Y \mid T=1,S=0]$. Essentially, it means that the average observed outcome in all experimental survivors is an estimate of average $Y(1)$ in \textit{Always-Survivors} diluted by presumably non-better outcomes of \textit{Experimental-only-Survivors}. The upper bound can be estimated from $n \times q$ best values of outcome $Y$ observed in the experimental group, with the proportion $q=p_0/p_1$, where $p_0$ and $p_1$ are proportions of survivors in the control and experimental groups, respectively. The upper bound estimate is reminiscent of a trimmed mean approach (Permutt and Li, 2017\cite{permutt2017}). See also a review of PS in Section 3 of Richardson et al. (2014) \cite{Richardson2014}.  

\section{Principal stratification based on joint modeling of continuous potential outcomes}\label{sc.partial.compliance}
In previous sections we considered principal strata based on potential outcomes of a discrete post-baseline variable, $S$. Here we briefly review applications where principal strata are based on POs that are inherently continuous. Like with PS based on discrete outcomes, the development was motivated by studying compliance to treatment, specifically when measured by a continuous variable (e.g., reflecting the proportion of pills taken by each patient during a certain period). Although here we only consider applications related to ``partial compliance,'' one could imagine modeling causal effects based on principal strata defined by other continuous outcomes.

Efron and Feldman (1991)\cite{Efron1991} (further referred to as EF) analyzed data from a double-blinded clinical trial where patients were randomized to receive a cholesterol lowering drug or placebo. In both groups compliance was not perfect resulting in most patients randomized to active treatment receiving only part of their full dose. EF used imperfect compliance as a natural basis for undertaking a causal analysis of dose effect trying to mimic a hypothetical trial where different patients would be randomized to specific doses while enforcing 100\% compliance to that dose. To facilitate inference, they assumed a certain deterministic relationship between the distribution of compliance in treated and control groups. As observed by EF, better compliance to drug was associated with larger reductions in cholesterol levels. However, there was also a positive trend in control group. This can be explained by a purely physiological component of compliance that may be indirectly related to cholesterol levels via correlation with a common latent variable such as propensity of compliers to eat healthier food.

Jin and Rubin (2008) \cite{jin2008principal} (hereafter, JR) reanalyzed the data by casting it within a principal stratification framework, which allowed them to use more precise causal language and specify more nuanced and plausible assumptions. To fix the ideas, $0 \le D_{0i}(0), D_{0i}(1) \le 1$ are two potential outcomes for partial compliances with respect to control treatment (indicated with subscript ``$0$'') for a patient $i$ if randomized to control and treatment arms, respectively. Similarly, $0 \le D_{1i}(0), D_{1i}(1) \le 1$ are partial compliances of the same patient with respect to the active treatment (indicated with subscript ``$1$''). As before, we will drop subject subscript unless it causes confusion. The interest is in estimating treatment effect within strata defined by various combinations of levels of these four variables, $S=(D_1(1),D_1(0),D_0(1),D_0(0))$.

With this general notation we can see that our four strata of ``Compliers'', ``Never-takers'', ``Always-takers'', and ``Defiers''defined in Table~\ref{tab:compl} of Section~\ref{sc.ps.trtchange} is a special case of strata defined by $S=(D_1(1),D_1(0),0,0)$, with principal strata formed by four combinations of \textit{binary} stratification variables, $D_1(0),D_1(1)\in \{0,1\}$. Alternatively, in Table~\ref{tab:compl2} of Section~\ref{sc.ps.monotonicity} we considered arbitrary patters of compliance with stratification based on $S=(D_1(1),0,0,D_0(0))$ with 4 principal strata ``Always-compliers'', ``Control-only-compliers'', ``Experimental-only-compliers'' and ``Never-compilers'' formed as combinations of binary stratification variables $D_0(0),D_1(1)\in \{0,1\}$.

JR re-analyzed the data from EF assuming so-called ``strong access monotonicity'' $D_1(0)=D_0(1)=0$ (patients randomized to treatment $t$ cannot have access to alternative treatment $1-t$) therefore their strata was based on $S=(D_1(1),0,0,D_0(0))$, where $D_1(1)$ and $D_0(0)$ unlike in the setting of Section~\ref{sc.ps.monotonicity} can assume any values from 0 to 1. The goal is in estimating the average causal treatment effect within the subset $S$ with $D_1(1)$ and $D_0(0)$ fixed at any combination of levels within the range from 0 to 1. The principal causal effect therefore is defined as $PCE(d_1,d_0)=E(Y(1)-Y(0)|D_1 (1)=d_1,D_0(0)=d_0)$. Note a connection with causal effect predictiveness (CEP) surface of Gilbert and Hudgens (2008) \cite{Gilbert2008} where conditioning is on arbitrary potential levels of a biomarker $S$ measured after treatment assignment, that is CEP is conditional on $(S(1)=s_1, S(0)=s_0)$. JR also assumed ``negative side-effect monotonicity,'' $D_1(1) \le D_0(0)$, which can be naturally interpreted when lack of compliance is caused by adverse effects associated with active treatment, as a result the same patient would have lower compliance when receiving the active drug than when receiving control. As usual, the SUTVA and treatment ignorability of PO`s (ensured by randomization to treatment) were made.

JR proposed a Bayesian parametric modeling of the joint distribution of partial compliances. They specified a beta distribution for $D_0(0) \sim Beta(\alpha_1,\alpha_2)$ and another beta distribution for relative drug compliance conditional on $D_0(0)$, $D_1(1)/D_0(0) \sim Beta(\alpha_3,\alpha_4)$. Constraining $D_1 (1)/D_0(0) \le 1$ is consistent with the negative side-effect monotonicity assumption. Potential outcomes $Y(1)$ and $Y(0)$ given both $D_1(1)$ and $D_0(0)$ were modeled using regressions with normal errors and linear effects for $D_1(1)$ and $D_0(0)$ for $Y(0)$ and additional quadratic effect in $D_1(1)$ for $Y(1)$ to accommodate evidence of a strong dose response in treated subjects. Bayesian estimation proceeded using basic ideas of data augmentation when missing potential outcomes are treated as missing data. An MCMC (Gibbs) sampler\cite{tannerwang1987DA} was used to draw from full conditional distributions of $D_1(1)$ and $D_0(0)$. Upon convergence of the MCMC, missing potential outcomes for $Y(1)$ and $Y(0)$ were drawn from their posterior distribution and individual treatment effects $Y(1)-Y(0)$ computed within each principal stratum. Importantly, $Y(1)$ and $Y(0)$ were assumed conditionally independent given principal strata $D_1(1)$ and $D_0(0)$, which corresponds to the assumption of principal ignorability. Clearly, observed data does not allow modeling conditional correlation in potential outcomes, $\rho=\text{corr}(Y(1),Y(0)|D_1(1),D_0(0))$. Therefore, JR proposed that $\rho$ be treated as a sensitivity parameter and principal effects for different combinations of $D_1(1)$ and $D_0(0)$ were estimated under varying assumed values of $\rho$. The authors found the changes in estimated PS effects were minimal after assuming non-zero correlations. Their general conclusion was that the drug effect on reduction in cholesterol levels was largest within  the strata of perfect compliers, that is for $S=(1,0,0,1)$. JR further conducted a more elaborate analysis of dose response within substrata of patients having the same placebo compliance, $D_0(0)$ using similar modeling tools under some additional assumptions (described in their section 4). 

Bartolucci and Grilli (2011)\cite{Bartolucci2011} (hereafter, BG) extended the model of JR in several ways and provided another set of reanalyzes of EF data set by modeling joint distribution of potential compliances $D_1(1)$ and $D_0(0)$ in the treated and control groups, respectively, through the Plackett copula thus avoiding any monotonicity assumptions inherent in parametric modeling by JR. Their analysis also differs from that by JR in allowing more flexibility in the potential outcome regressions such as including interaction terms with $D_1(1)$ by $D_0(0)$ and heteroscedastic errors. The copula allowed them to study the association between the latent compliances, $D_1(1)$ and $D_0(0)$, without specifying parametric models for their marginal distributions, which were estimated by their empirical distribution functions from observed data in the experimental and control arms, respectively. Similar approaches utilizing copula for describing relationship between potential outcomes were used in mediation analysis via principal effects\cite{Kim2019}. The joint distribution of compliances is governed by association parameter $\psi$ (with $\psi=1$ indicating independence), which has a simple connection with the Spearman's rank correlation. For each value of $\psi$, joint distribution of compliances is estimated using Plackett copula; then $Y(0)$ and $Y(1)$ are modeled as regression functions of  $D_0(0)$ and $D_1(1)$, estimated via EM algorithm for maximum likelihood. Like BG, they assumed conditional indepednece of $Y(1)$ and $Y(0)$, given a stratum. Although $D_0(0)$ and $D_1(1)$ are never jointly observed, the missing compliances can be integrated from the joint likelihood of $Y(0)$ and $Y(1)$ because the conditional distributions of $D_1(1)|D_0(0)$ and $D_0(0)|D_1(1)$ can be obtained through the copula. BG reported point estimates and bootstrap based confidence intervals for parameters of the final selected model. As shown in BG, the association parameter $\psi$ can be estimated using profile likelihood, however, they warn that the empirical support for profile ML is rather scarce and there are many values of $\psi$ that may be equally well supported by the data as indicated by flat regions of profile likelihood function near the maximum. Therefore, adopting a sensitivity framework by conducting analyses for a set of values of $\psi$ within plausible regions is preferred.

Evaluating principal causal effects under partial compliance is further motivated by complex multistage sequential multiple assignment randomized trials (SMART) where compliance during different stages is often defined as the average compliance measured through the follow-up time which is a continuous variable. Here interest may be in evaluating effects conditional on specific partial compliances at various stages. For example, Artman et al. (2020)\cite{Artman2020} and Bhattacharya et al. (2021)\cite{Bhattacharya2021} proposed a Bayesian semiparametric approach for estimating the mean treatment strategy outcome given compliance classes (i.e., there is a stratum/class for any combination of potential partial compliances at different stages of the trial). They use a semi-parametric Bayesian model where the joint distribution of compliances for treated and control subjects are estimated using a Gaussian copula and Dirichlet process mixture is used for modeling potential outcomes.

\section{Estimators using baseline covariates}\label{sc.ps.baseline}

One way to estimate outcomes in principal strata is by employing information contained in baseline covariates $X$ while making a (rather strong) assumption that given $X$, strata membership would provide no additional information for predicting potential outcomes. Consider a setting described in Section \ref{sc.ps.outcomes} with principal stratification as represented in Table~\ref{tab:symptom}, where the interest is in estimating the treatment effect for a  continuous outcome (extension to binary outcomes is straightforward) in the \textit{Immune} stratum. Here, the monotonicity assumption implies that the \textit{Harmed} stratum is empty, the distribution of outcomes under the control treatment,  $F_0(y \mid I) = \textrm{Pr(}Y(0) \le y \mid I)$ can be estimated from the observed data in the control arm from patients with $S(0)=0$. The difficulty is in identifying the distribution of outcome under the experimental treatment in the \textit{Immune} stratum, $F_1(y \mid I)=\textrm{Pr(}Y(1) \le y \mid I)$ because $S(0)$ is unobserved for patients in the experimental arm:
\[
F_1(y|I)=\int_{-\infty}^{y} f_{Y(1)}(u\mid S(0)=0)du.
\]
We use the information contained in baseline covariates $X$ assuming that knowledge of $S(0)$ provides no additional information for predicting the potential outcome $Y(1)$ after $X$ has been fitted. This can be expressed as $(S(0) \ind Y(1)|X)$.

Following Bornkamp and Bermann (2019) \cite{bornkamp2019estimating}, we rewrite $F_1(y \mid I)$ in two different forms that give rise to two approaches for estimating the treatment effect in principal strata: based on ($i$) predicted counterfactual response or ($ii$) weighting by propensity of strata in the control arm. The latter approach is closely related to the principal score based methods of Section~\ref{sc.ps.baseline.principal.scores}. These ideas are presented in the next two subsections.

\subsection{Predicted counterfactual response}\label{sc.ps.baseline.cf.response}
Assuming conditional independence of $Y(1)$ and $S(0)$, given $X$, we can express the density for treated, conditional on the \textit{Immune} stratum as 
\begin{eqnarray}
\nonumber
f_1(y \mid I) &=& f_{Y(1)}(y \mid S(0)=0) \\ &=&  
\nonumber \int_x f_{Y(1)}(y \mid S(0)=0,X=x)f(x \mid S(0)=0)dx \\ &=&
\nonumber \int_x f_{Y(1)}(y \mid X=x)f(x \mid S(0)=0)dx \\ &=&
\nonumber \int_x f_{Y}(y \mid X=x, T=1)f(x \mid S=0, T=0)dx. 
\end{eqnarray}
Note that in the last line we used independence of potential outcomes $Y(0), Y(1)$ of actual treatment assignment $T$ (under randomization) as well as the consistency assumption (as part of SUTVA) that under treatment $T=t$ the observed outcome $Y$ is the same as potential outcome $Y(t)$.

Using a general form of the test statistic (\ref{eq.gen.stat}) (see Section~\ref{sc.ps.outcomes.continuous}), the first component of the difference can be estimated as a simple average of the outcomes for patients randomized to control arm $\{T=0\}$ who had no event $\{S=0\}$. The second component can be evaluated using predicted response for patients in the same subset $\{T=0,S=0\}$, if they were under experimental treatment (contrary to the fact). This suggests the following test statistic 
\begin{equation}\label{eq.pred.count.resp}
T_1=\frac{1}{n_0} \sum_{i=1}^N (1-t_i)(1-s_i)(y_i-\hat{m}_1(x_i)),
\end{equation}
where $\hat{m}_1(x_i)$ is a "predicted counterfactual response" based on a regression of Y on X estimated from all treated patients, which is evaluated for covariate profile $x_i$ for each control patient in group $S=0$. 

Here, unlike Bornkamp and Bermann (2019) \cite{bornkamp2019estimating}, we are predicting the response in control patients if randomized to experimental treatment rather than predicting response in the experimental arm if randomized to control. Hence, the method is labeled as ``predicted placebo response'' in \cite{bornkamp2019estimating}, while  in our case it would be ``predicted treated response.'' To generalize we labeled it as ``predicted counterfactual response'' to emphasize that the method requires predicting response that would have been observed under treatment different than the one assigned at randomization. 

Louizo et al. (2017) \cite{louizos2017causal} proposed building the prediction model for counterfactual response under treatment $T=t$ via intermediate outcomes $Z$ using the patients randomized to $T=t$ and then conditioning on the baseline covariates. That is, the mean function $m_t$ can be estimated as
\[
m_t(x) = E\{E(Y|Z,X,T=t)|X=x\}.
\]
Here the inner expectation is taken with respect to $Z$ and the outer with respect to $Y$. The use of intermediate outcomes in the above double expectation has two advantages: (1) it may provide a more robust prediction function, especially if $X$, $Z$ and $Y$ are not from a multivariate normal distribution, and (2) it fully utilizes the repeated measurements. 

\subsection{Strata propensity weighted estimator}\label{sc.ps.baseline.prop.weighted}
The expression for $f_1(y|I)$  can be re-written as 
\begin{eqnarray}
\nonumber
f_1(y \mid I) &=& f_{Y(1)}(y \mid S(0)=0) \\ &=&  
\nonumber \int_x f_{Y(1)}(y \mid S(0)=0,X=x)f(x \mid S(0)=0)dx \\ &=&
\nonumber \int_x \frac{f_{Y(1)}(y \mid X=x)\textrm{Pr(}S(0)=0 \mid X=x)f(x)}{\textrm{Pr(}S(0)=0)}dx \\ &=&
\nonumber \int_x \frac{f_{Y}(y \mid X=x, T=1)\textrm{Pr(}S(0)=0 \mid X=x, T=0)f(x)}{\textrm{Pr(}S(0)=0 \mid T=0)}dx \\ &=&
\nonumber \int_x \frac{f_{Y}(y \mid X=x, T=1)\textrm{Pr(}S=0 \mid X=x, T=0)f(x|T=1)}{\textrm{Pr(}S=0 \mid T=0)}dx. 
\end{eqnarray}
Note that in the last line we replaced $f(x)$ with $f(x|T=1)$ by randomization. 

By letting $w_0(x)=\textrm{Pr(}S=0 \mid X=x,T=0)$, the probability of observing $S=0$, given covariates and $T=0$, we can re-write as
\[
f_1(y \mid I)=\frac{1}{\textrm{Pr(}S=0 \mid T=0)} \int_x f_{Y}(y \mid X=x, T=1)w_0(x)f(x|T=1)dx.
\]
The weight function $w_0(x)$ can be estimated using logistic regression. Note that $w_0(x)$ is closely related to principal scores, which are discussed in the next section. Now we can estimate outcomes in treated patients for the \textit{Immune} stratum as
\[
F_1(y|I)=\frac{1}{\textrm{Pr(}S=0 \mid T=0)} \int_{-\infty}^{y} \int_x f_{Y}(u \mid X=x, T=1)w_0(x)f(x|T=1)dxdu.
\]
Consequently, a test statistic can be constructed as
\begin{equation}\label{eq.ps.weight}
T_2=\frac{1}{n_0} \sum_{i=1}^N (1-t_i)(1-s_i)y_i- \frac{1}{N_1}\frac{1}{\widehat{Pr}(S=0 \mid T=0)} \sum_{i=1}^{N} t_i\hat{w}_0(x_i)y_i,
\end{equation}
where $N_1$ is the number of patients in the treatment arm, and $\hat{w}_0(x_i)$ is the estimated probability of $S=0$, based on a logistic regression fitted to the control arm and evaluated on a patient in experimental arm with a covariate profile $x_i$.

We do not need the monotonicity assumption if instead of assuming independence $S(t) \ind Y(1-t)|X$, we assume $S(t) \ind S(1-t) \mid X$, $t=0,1$ (see also Hayden, Pauler, and Schoenfeld, 2005 \cite{hayden2005estimator}). The plausibility of these assumptions is discussed in Section~\ref{sc.summary}. 
\begin{eqnarray}
\nonumber
f_1(y \mid I) &=& f_{Y(1)}(y \mid S(0)=0, S(1)=0) \\ &=&  
\nonumber \int_x f_{Y(1)}(y \mid S(0)=0, S(1)=0, X=x)f(x \mid S(0)=0, S(1)=0)dx \\ &=&
\nonumber \int_x \frac{f_{Y(1)}(y \mid X=x)\textrm{Pr(}S(0)=0,S(1)=0 \mid X=x)f(x)}{\textrm{Pr(}S(0)=0, S(1)=0)}dx \\ &=&
\nonumber  \frac{\int_x f_{Y(1)}(y \mid X=x)\textrm{Pr(}S(0)=0 \mid X=x)\textrm{Pr(}S(1)=0 \mid X=x)f(x)dx}{\int_x \textrm{Pr(}S(0)=0 \mid X=x)\textrm{Pr(}S(1)=0|X=x)f(x)dx} \\ &=&
\nonumber  \frac{\int_x f_{Y}(y \mid X=x, T=1)\textrm{Pr(}S=0 \mid X=x, T=0)\textrm{Pr(}S=0 \mid X=x, T=1)f(x \mid T=1)dx}{\int_x \textrm{Pr(}S=0 \mid X=x, T=0)\textrm{Pr(}S=0 \mid X=x, T=1)f(x\mid T=1)dx}. 
\end{eqnarray}
All the quantities in the last line can be estimated from the observed data. 

Similarly, we can express $f_0(y \mid I)$ as 
\begin{eqnarray}
\nonumber
f_0(y \mid I) &=& f_{Y(0)}(y \mid S(0)=0, S(1)=0) \\ &=&  
\nonumber  \frac{\int_x f_{Y}(y \mid X=x, T=0)\textrm{Pr(}S=0 \mid X=x, T=0)\textrm{Pr(}S=0 \mid X=x, T=1)f(x \mid T=0)dx}{\int_x \textrm{Pr(}S=0 \mid X=x, T=0)\textrm{Pr(}S=0 \mid X=x, T=1)f(x\mid T=0)dx}. 
\end{eqnarray}

Consequently, a test statistic for the null hypothesis of treatment effect (like in \eqref{eq.gen.stat}), can be constructed as 
\begin{equation}\label{eq.base.nomon1}
T_3=\frac{\sum_{i=1}^{N} (1-t_i)(1-s_i)\hat{w}_1(x_i)y_i}{\sum_{i=1}^{N} (1-t_i)(1-s_i)\hat{w}_1(x_i)} - \frac{\sum_{i=1}^{N} t_i(1-s_i)\hat{w}_0(x_i)y_i}{\sum_{i=1}^{N} t_i(1-s_i)\hat{w}_0(x_i)},
\end{equation}
where $w_t(x)=\textrm{Pr(}S=0 \mid X=x,T=t), t \in \{0,1\}$.
While (\ref{eq.base.nomon1}) does not require a monotonicity assumption, as other estimators for PS it requires unverifiable assumptions. Note that in the first term of the right-hand side of expression (\ref{eq.base.nomon1}), we replaced the  probability of strata membership in control group, $\textrm{Pr}(S=0|X=x,T=0)$, with the observed indicators ($1-s_i$); similarly, for the treated group in the second term of (\ref{eq.base.nomon1}). This was done to minimize the amount of modeling and use observed data as much as possible. Alternatively, we could use an estimator involving both $\hat{w}_0(x)$ and $\hat{w}_1(x)$ for estimating the mean response for each treatment arm in the \textit{Immune} stratum:
\begin{equation}\label{eq.base.nomon2}
T_4=\frac{\sum_{i=1}^{N} (1-t_i)\hat{w}_1(x_i)\hat{w}_0(x_i)y_i}{\sum_{i=1}^{N} (1-t_i)\hat{w}_1(x_i)\hat{w}_0(x_i)}- \frac{\sum_{i=1}^{N} t_i\hat{w}_1(x_i)\hat{w}_0(x_i)y_i}{\sum_{i=1}^{N} t_i\hat{w}_1(x_i)\hat{w}_0(x_i)}.
\end{equation}

Selection between (\ref{eq.base.nomon1}) and (\ref{eq.base.nomon2}) can be driven by the bias-variance trade-off. One can argue that while (\ref{eq.base.nomon1}) may have larger variance, as it is based on a subset of patients with $S=0$, it may be more robust to model misspecifications in estimating $w_t(x)$ (resulting in smaller bias). 

\subsection{Methods based on principal scores }\label{sc.ps.baseline.principal.scores}

Principal scores are similar to propensity scores for estimating treatment effects in observational studies with a non-random treatment assignment (see Jo and Stuart, 2009 \cite{jo2009use};  Ding and Lu, 2017 \cite{ding2017principal}; Feller et al., 2017 \cite{feller2017principal}). Like propensity scores, the principal score is a balancing score in that the distribution of covariates is similar within principal strata conditional on the principal score. Our discussion follows that in Ding and Lu\cite{ding2017principal}. 

Similar to the approach of Bornkamp and Bermann (2020) \cite{bornkamp2019estimating}, we use the strong version of the Principal Ignorability assumption discussed in Section \ref{common.assumptions}:
\[
Y(1) \ind (S(0),S(1))\mid X \; and \; Y(0) \ind (S(0),S(1)) \mid  X.
\]
In other words, under strong PI, given covariates, stratum membership can be considered as if assigned ``at random'' and we can equate conditional expectations of $Y$ across strata for $t=0,1$:
\begin{eqnarray}
\nonumber E[Y(t) \mid X, S(0)=0, S(1)=0] &=& \\
\nonumber E[Y(t) \mid X, S(0)=0, S(1)=1] &=& \\
\nonumber E[Y(t) \mid X, S(0)=1, S(1)=0] &=& \\
\nonumber E[Y(t) \mid X, S(0)=1, S(1)=1]. 
\end{eqnarray}
With a large number of covariates, an attractive option is to summarize the dependency of strata membership on covariates in a single-dimensional score that can be used in place of the original $p$-dimensional covariate vector $x$. This is analogous to propensity scores used as balancing scores in the analysis of non-randomized experiments (Jo and Stuart, 2009 \cite{jo2009use}). 

Let the four principal strata be denoted with a single multinomial variable,
\begin{equation}\label{eq.prstrata}
 U=\{S_{00},S_{01},S_{10},S_{11}\}, S_{ij}=(S(0)=i,S(1)=j), i,j=0,1.
\end{equation}
Using the example of Section~\ref{sc.ps.outcomes} with four strata $S_{00},S_{01},S_{10},S_{11}$ representing \textit{Immune}, \textit{Harmed}, \textit{Benefiters}, and \textit{Doomed}, respectively, as summarized in Table~\ref{tab:symptom}. Define the \textit{principal score} as
\[
\pi_u(x)=\textrm{Pr(}U=u \mid X=x).
\]
Under the strong PI, the principal score enjoys a balancing property: $U \ind X \mid \pi_u(x)$. This is similar to the balancing property of the propensity score in observational studies: $ T \ind X \mid p(x)$, where $p(x)= \textrm{Pr(}T=1 \mid X=x)$.

Principal scores can be estimated from observed data, under certain assumptions. For example, under monotonicity, implying $\textrm{Pr}(U=S_{01})=0$ (i.e., zero chance of being harmed), the probabilities in the three remaining cells of the multinomial $U$ can be estimated as functions of covariates using the following ideas. 

A na\"{\i}ve strategy is to go in stages. First, estimate (e.g., using logistic regression with various baseline characteristics $X$ as covariates) the probability of \textit{Immune} stratum membership given patient's covariates $X$ from the control arm alone, $\textrm{Pr(}U=S_{00} \mid X=x)=\textrm{Pr(}S=0 \mid X=x,T=0)$,  
\[
\hat{\pi}_{00}(x)=\widehat{\textrm{Pr}}(S=0 \mid X=x,T=0).
\]
Then find remaining probabilities $\pi_{11}(x)$ and $\pi_{10}(x)$ by subtraction from the marginal probabilities. E.g.,  
\[
\hat{\pi}_{10}(x)=\widehat{\textrm{Pr}}(S(1)=0 \mid X=x)- \hat{\pi}_{00}(x).
\]
However, this may lead to inconsistent results in that if we start with estimating the probability of \textit{Doomed}, as $\textrm{Pr(}U=S_{11} \mid X=x)=P(S=1 \mid X=x,T=1)$: 
\[
\hat{\pi}_{11}(x)=\widehat{\textrm{Pr}}(S=1 \mid X=x,T=1),
\]
and then determine the two other probabilities by subtraction, we may get different results.

A more robust approach is to simultaneously estimate probabilities in all three non-empty cells using a mixture approach treating unknown stratum as missing data or a latent variable via the well-known Expectation Maximization (EM) or MCMC algorithms\cite{ding2017principal}. Once the probabilities of strata membership are estimated, we can construct principal strata estimators as follows. Assume the estimation target is $\delta=E(Y(1) \mid I)-E(Y(0) \mid I)$ and the main challenge is identifying the first expectation which under monotonicity is $E(Y(1) \mid S(0)=0)$. This can be expressed via observable outcomes using principal scores\cite{ding2017principal}.  
\begin{eqnarray}
\nonumber
\nonumber E(Y(1) \mid S(0)=0)  &=&
\nonumber E(Y(1) \mid S(0)=0, T=1, S=0)  \\ &=&
\nonumber E_X \{E\left(Y(1) \mid S(0)=0, T=1, S=0, X=x\right)\} \\ &=& 
\nonumber \int_x  E(Y(1) \mid S(0)=0, T=1, S=0, X=x) f(x\mid S(0)=0,T=1, S=0)dx  \\ &=&
\nonumber \int_x  E(Y(1) \mid S(1)=0, T=1, S=0, X=x) f(x\mid S(0)=0,T=1, S=0)dx  \\ &=&
\nonumber \int_x  E(Y \mid T=1, S=0, X=x) \frac{\textrm{Pr(}S(0)=0 \mid X=x, T=1, S=0)}{\textrm{Pr(}S(0)=0 \mid T=1, S=0)} f(x\mid T=1, S=0)dx  \\ &=&
\nonumber \int_x  E(Y \mid T=1, S=0, X=x) \frac{\frac{\pi_{00}(x)}{\pi_{00}(x)+\pi_{10}(x)}}{\frac{\pi_{00}}{\pi_{00}+\pi_{10}}} f(x\mid T=1, S=0)dx.
\end{eqnarray}

The first line uses the fact that under randomization $Y(1)$ is independent of treatment assignment $T$ and that under monotonicity (no one is ``harmed'') $(S(0)=0)$ implies $S(1)=0$, which within the treated arm $T=1$ implies $(S=0)$. The second line uses the law of iterated expectations. The fourth line uses the principal ignorability assumption which allows replacing $(S(0)=0)$ with $(S(1)=0)$ in the conditioning part. As a result, we can replace (in the fifth line) the potential outcome $Y(1)$ with observable $Y$ (under SUTVA). Also, we use the Bayes rule in line five to re-express the $X$-density conditional on $\{S(0)=0\}$ via the probabilities of strata membership and the density unconditional on $S(0)$. 

The equivalence $\textrm{Pr(}S(0)=0 | X=x, T=1, S=0)=\frac{\pi_{00}(x)}{\pi_{00}(x)+\pi_{10}(x)}$ allows us (in the last line) to express the conditional probability of being in the \textit{Immune} stratum via the ratio of principal scores for the \textit{Immune} to the sum of the principal scores in the \textit{Immune} and \textit{Benefiters}. This can be easily shown using monotonicity and randomization assumptions. 

Finally denoting the subject weight as
\[
w_{00}(x)=\frac{\frac{\pi_{00}(x)}{\pi_{00}(x)+\pi_{10}(x)}}{\frac{\pi_{00}}{\pi_{00}+\pi_{10}}},
\]
we can express the unknown quantity via observables and weights computed from estimated principal scores.
\[
E(Y(1) \mid S(0)=0) = \int_x E(Y \mid T=1, S=0, X=x)w_{00}(x)f(x\mid T=1, S=0)dx.
\]
A non-parametric method of moments estimator for $\delta$ can be constructed by replacing integration with summations over observed outcomes within the group $\{T=1,S=0\}$.
\begin{equation}\label{eq.pr.score.est}
\widehat{\delta}= \frac{\sum_{i:t_i=1,s_i=0} \widehat{w}_{00}(x_i)y_i}{\sum_{i:t_i=1,s_i=0}\widehat{w}_{00}(x_i)} -\frac{\sum_{i:t_i=0,s_i=0} y_i}{\sum_{i=1}^N {I(t_i=0, s_i=0)}},
\end{equation}
where $I(\cdot)$ is the indicator function. After incorporating baseline covariates into the model for strata membership, these covariates can also be used to better model the outcomes $Y$ using parametric and semiparametric modeling.

Sensitivity analyses to the PI assumption are necessary. It is tempting to limit such analyses to relatively benign cases of model misspecification, such as a misspecified link function (logit vs. probit) or assuming omitted covariates for principal score modeling (see, e.g., Shen, Ning and Yuan, 2015 \cite{shen2015bayes}). However, it is important to also conduct sensitivity analyses that challenge the PI assumption by assuming that the omitted covariates are related to the future potential outcomes $Y(t)$. This is similar to sensitivity analysis for propensity score methods in observational data where we should not only assume that our propensity model may have missed some important predictors but that these predictors are related with outcomes (i.e., the omitted variables are unmeasured confounders). For sensitivity analyses in the context of principal scores see Ding and Lu\cite{ding2017principal}.

\subsection{Implementing principal stratification strategy via multiple imputation}\label{sc.ps.imputation}

An attractive strategy for evaluating various estimands based on principal stratification is multiple imputation (MI). Since principal strata membership (as any potential outcome) can be formulated as a missing data problem, various multiple imputation strategies can be implemented. Here missing data corresponds to the unknown post-randomization event $S(1)$ for $T=0$ and $S(0)$ for $T=1$. As before, we assume that baseline covariates contain useful information for estimating principal strata membership. Therefore, one imputation strategy is simply to impute missing $S(t)$ for patients in treatment arm $T=1-t$ using baseline covariates alone and then proceed with estimating the treatment effect in any principal stratum by sub-setting each of the $m$ completed data sets on the stratum of interest (observed or imputed) and applying standard estimation procedures to the observed outcomes $Y$ within that subset for each completed set (see a more detailed description of this strategy with a case study in Chapter 26 of Mallinckrodt et al. 2020 \cite{mallinckrodt2020}). If missing outcomes $Y$ are encountered within the stratum of interest, these can be imputed using an additional imputation model following the imputation of $S(t), t=0,1$ or dealt with using maximum likelihood methods for repeated measures. The final estimate of treatment effect is computed by combining estimates of the stratum-specific treatment effects across the $m$ completed sets using Rubin's rules\cite{rubin1987}. The data setup corresponding to this strategy is presented in Table~\ref{table_MI} with missing data being represented either by counterfactuals or ``true'' missing values.  

As an illustration of this approach, we apply it to computing the treatment effect in a stratum of patients who would be adherent to either experimental drug or control for our diabetes example of Section~\ref{data.example}. That is, like in Section~\ref{sc.ps.outcomes.continuous.example} we are interested in \textit{Always-compliers}, where the stratum of interest $S_{00}=\{S(0)=0,S(1)=0\}$. The imputation model for missing counterfactual compliance  $S(t),t=0,1$ is a Bayesian logistic regression with  age, gender, baseline HbA1c, LDL-C, triglycerides, fasting serum glucose, and alanine aminotransferase as covariates\cite{qu2020imp}. As a result, $m=100$ sets with completed potential outcomes $S(0), S(1)$ for each patient were constructed. Note that all patients in the substratum $S_{00}$ have non-missing outcome $Y$, therefore estimators of treatment effect within the stratum can be obtained by a simple ANCOVA model for HbA1c at the last evaluation visit with the treatment indicator and baseline HbA1c as covariates. On average, 62.6\% of all patients fell within stratum $S_{00}$ across 100 imputed data sets. The resulting point estimate and 95\% CI constructed by using Rubin's rules were $-0.244$ and $(-0.36; -0.13)$, respectively. 

Note that this approach of imputing missing strata membership alone would not allow us to estimate treatment effect in stratum $\{S(1)=0\}$ comprised of patients who would comply with the experimental arm regardless of randomized treatment. That is because we need to also impute missing outcomes for those control patients in this stratum with intercurrent event ($S=1$), which could be done by incorporating in the imputation model repeated measures of HbA1c. 

\begin{table}[h!tb] \centering
\caption{Illustration of data setup for imputing missing potential outcomes $S(t)$ under treatment $T=t \in \{0,1\}$ for patients assigned to treatment $T=1-t$. $S=1$ indicates intercurrent event\textemdash in this example being the only cause for unobserved outcome $Y$} 
\begin{tabular}{ccccccc}
\hline
Subject & Randomized treatment ($T$) & $X$ & $S(0)$ & $S(1)$ & $Y$ \\
\addlinespace[0.2cm]
\hline
001 & $0$ & $\checkmark$ & 0 & $\cdot$ & $\checkmark$ \\
002 & $0$ & $\checkmark$ & 0 & $\cdot$ & $\checkmark$ \\
003 & $0$ & $\checkmark$ & 1 & $\cdot$ & $\cdot$ \\
$\cdots$ &&&&&&\\
101 & $1$ & $\checkmark$ & $\cdot$ & 1 & $\cdot$ \\
102 & $1$ & $\checkmark$ & $\cdot$ & 0 & $\checkmark$ \\
103 & $1$ & $\checkmark$ & $\cdot$ & 0 & $\checkmark$ \\
\hline
\end{tabular}
{\footnotesize \begin{flushleft} \textbf{Abbreviations}: ``$\checkmark$'', non-missing data; ``$\cdot$'', missing data. \end{flushleft}}
\label{table_MI}
\end{table}

A more advanced approach is to impute jointly missing potential outcomes for both principal stratum variable $S$ and outcome variable $Y$ taking advantage of repeated measures for $Y$, the exact timing of the intercurrent event of interest $S$, and possibly other intermediate variables $Z$ as early predictors of both $S$ and $Y$. For convenience we can include repeated measures of $Y$ in a vector of intermediate covariates $Z$. The imputation strategy is outlined in Table \ref{table_MI_ext} in Section~\ref{sc.ps.baseline.post-baseline} with more details provided in Luo et al. (2021) \cite{luo2020mi}. 

In this setup, imputation strategies appropriate for  monotone patterns of missing data can be employed allowing direct sampling from posterior predictive distributions of missing data given observed data (using Bayesian regressions for modeling $Y|S,X$ and $S|X$). As usual, the analysis consists of three steps. First, $m$ completed data sets are produced via imputation. Second, estimates of the treatment effect within the principal strata of interest are computed from each completed set. Third, the point estimates and standard errors from the $m$ sets are combined in a single point estimate and confidence interval using either Rubin's rules or bootstrap. 

\section{Estimators using baseline and post-baseline covariates}\label{sc.ps.baseline.post-baseline}
Louizos et al. \cite{louizos2017causal} proposes methods directly estimating potential outcomes of the response variable if the principal stratum can be observed. Qu et al. \cite{qu2020general} provides a more general framework by modeling the potential outcomes of the response variable and/or the principal score via baseline covariates and potential post-baseline intermediate measurements for principal stratification defined by  treatment adherence status. In this section, we summarize the more general framework of Qu et al. \cite{qu2020general}. 

Assume that the stratification outcome $S$ denotes the indicator for the presence of intercurrent event with $S=0$ indicating patients adhere to the treatment without intercurrent events. Note in the original research, Qu et al. used $A$ to denote adherence status with $A=1$ indicating adherers, so $S = 1 - A$. In this study, a patient is adherent to the assigned treatment if they continue taking the assigned treatment through the planned study treatment period. 

The approach relies on several assumptions. In addition to the SUTVA assumptions (A1-A3) introduced in Section \ref{background}, we also require the following assumptions:

\begin{eqnarray}
\mbox{A4}: && T_i \ind \{Y_i(0), Y_i(1), S_i(0), S_i(1), Z_i(0), Z_i(1) \} | X_i \nonumber \\
\mbox{A5}: && S_i(t) \ind \{Y_i(1), Y_i(0), Z_i(1-t)\} | \{X_i, Z_i(t)\}, \quad \forall \; t=0,1 \nonumber \\
\mbox{A6}: && Y_i(t) \ind Z_i(1-t) | \{X_i, Z_i(t)\}, \quad \forall \; t=0,1
\nonumber \\
\mbox{A7}: && Z_i(0) \ind Z_i(1)|X_i \nonumber. 
\end{eqnarray}
A4 is the \textit{treatment ignorability} assumption and A5 is the \textit{principal ignorability} (or  \textit{adherence ignorability}) assumptions in Section~\ref{common.assumptions}. A6 and A7 are the cross-world independence assumptions conditional on baseline and/or postbaseline covariates. Assumption A6 and A7 are particularly strong, similar to (\ref{eq.crossworld.ind}), as discussed in Section~\ref{common.assumptions}.

\begin{figure}[h!tb] \centering
\setlength{\unitlength}{1.4cm}
\begin{center}
\begin{picture}(5, 1.5)
\put(0,0.2){{$T$}}
\put(2,1.0){{$X$}}
\put(2,-0.6){{$Z$}}
\put(4,1.0){{$S$}}
\put(4,0.2){{$Y$}}
\put(0.3, 0.3){\vector(1,0){3.5}}
\put(2.3, 1.1){\vector(1,0){1.5}}
\put(2.3, 1.1){\vector(2,-1){1.5}}
\put(2.3, -0.5){\vector(1,1){1.5}}
\put(2.3, -0.5){\vector(2,1){1.5}}
\put(2.1, 0.9){\vector(0,-1){1.2}}
\put(0.3, 0.3){\vector(2,-1){1.4}}
\end{picture}
\end{center}
\vspace{0.2in}
\caption{ Causal diagram showing the dependencies between treatment ($T$), baseline covariate ($X$), post-baseline intermediate variable ($Z$), intercurrent event indicator ($S$) and outcome ($Y$). This figure is adapted from Figure 1 in Reference \cite{qu2020general}.}
\label{plot:relationship}
\end{figure}
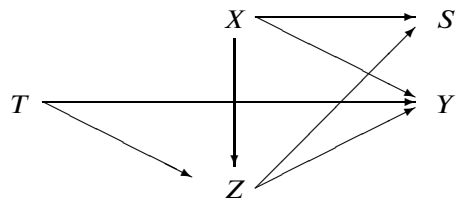
The relationships among $X_i, Z_i, Y_i$, and $S_i$ are described in the diagram in Figure \ref{plot:relationship}. We illustrate the methods for two principal strata: $S_{*0}=\{i:S_i(1)=0\}$ (patients that would adhere to the experimental treatment regardless of adherence to the control treatment) and $S_{00}=\{i:S_i(0)=0,S_i(1)=0\}$ (patients that would adhere to both treatments). Qu et al. \cite{qu2020general} suggested that $S_{*0}$ would be more relevant for placebo-controlled studies and $S_{00}$ be more relevant for active-comparator studies. For each set, the authors proposed two methods. 

Method A is based on the prediction of $Y_i(0)$ for patients who are randomized to the experimental treatment. The average treatment effect within the $S_{*0}$ stratum can be estimated as follows: 
\begin{equation} \label{eq:mu_d_s01_A}
 \widehat{E}[Y(1)-Y(0)|S_{*0}] = 
\frac{1}{n_{10}}\sum_{i \in \{i:t_i=1, s_i=0\}} y_i - \frac{1}{n_{10}} \sum_{i \in \{i:t_i=1, s_i=0\}} \hat \phi_0(x_i),
\end{equation}
where $n_{10}$ is the number of patients who adhere to the assigned experimental treatment and $\phi_0(X)$ is explained in the following. This estimator is also provided in Reference \cite{louizos2017causal}. The estimation steps for Method A are as follows:
\begin{enumerate}
\item Estimate the distribution of $Z_i(0)$ given $X_i$, denoted by $\hat F_{Z(0)|X}$, by using data from the control group, i.e., $\{(X_i, Z_i): i\in \{i:T_i=0\}\}$. As a common scenario, when both $X_i$ and $Z_i$ are continuous variables and can be assumed from normal distributions, then $F_{Z(0)|X}$ can be estimated using the conditional distribution formula for the bivariate (or multivariate) normal distribution. 
\item Build a prediction model for $Y$ using data from the control group i.e., $\{(Y_i, X_i, Z_i): i\in \{i:T_i=0\}\}$, denoted by $\hat\psi_0(X,Z(0))=\hat{E}[Y(0)|X,Z(0)]$. The observed data contains outcomes for patients only prior to ICE, so this requires A5 to provide valid inference. It is important to understand that $Y(0)$ here designates hypothetical outcomes that would have been observed had all patients from control arm been  compliant with their assigned treatment (even if contrary to the fact). Therefore, building the predictive model requires modeling repeated measures of $Y$. 
\item  Compute the conditional expectation $\hat \phi_0(X) =\hat{E}\{\hat\psi_0(X,Z(0))|X\}$ integrating over conditional distribution of intermediate covariates $Z|X$ using the distribution $\hat F_{Z(0)|X}$ from Step 1 and the prediction model $\hat \psi_0(X)$ from Step 2. In general, numerical integration is required for computing this expectation. If $X_i$, $Y_i$, and $Z_i$ are from a joint multivariate distribution, the multivariate normal distribution can be estimated using observed data $\{(Y_i, X_i, Z_i): i\in \{i:T_i=0\}\}$ and the conditional expectation $\hat \phi_0(X)$ can be directly computed from $E[Y(0)|X]$. 
\item Compute the estimated treatment effect using equation (\ref{eq:mu_d_s01_A}), where  $\hat \phi_0(X_i)$ is the estimated mean potential outcome under the control treatment for patients that adhere to experimental treatment group. In this step, Assumptions A4* and A5* are required.
\end{enumerate}

Method B is based on the prediction of $S_i(1)$ for patients who are randomized to the control treatment. The average treatment effect in the $S_{*0}$ stratum can be estimated via the functions of baseline and intermediate outcomes $h(X)$ and $g(X,Z)$ as explained in the following: 
\begin{equation} \label{eq:mu_d_s01_B}
 \widehat{E}[Y(1)-Y(0)|S_{*0}] = \frac{1}{n_{10}} \sum_{i \in \{i:t_i=1, s_i=0\}} y_i - 
\frac{n_1}{n_{10}n_0} \sum_{i \in \{i:t_i=0, s_i=0\}} \frac{\hat h_1(x_i) y_i}{\hat g(x_i,z_i)},
\end{equation}
where $n_t$ is the number of patients assigned to treatment $t$ and $n_{t0}$ is the number of patients adherent to treatment $t$. The estimation steps for Method B are as follows:
\begin{enumerate}
\item Estimate the distribution of $Z_i(1)$ given $X_i$ , denoted by $\hat F_{Z(1)|X}$, using data from the experimental treatment group, i.e., $\{(X_i, Z_i): i\in \{i:T_i=1\}\}$.  
\item Build a prediction model for treatment adherence conditional on the baseline covariates and observed intermediate outcomes using data from both treatment groups. The estimated prediction model is denoted by $\hat g(X,Z):=\widehat{\text{Pr}}(S=0|X,Z)$. Generally, a logistic regression model can be used to estimate the above conditional distribution. We assume adherence status does not depend on treatment given observed early outcomes $Z$. Mathematically, it means an additional treatment ignorability assumption: $S_i \ind T_i | X_i, Z_i$. Alternatively, we could use only patients from the control arm $\{(S_i, X_i, Z_i): i\in \{i:T_i=0\}\}$ to estimate $\hat g_0(X,Z):=\widehat{\text{Pr}}(S=0|X,Z,T=0)$.
\item  Compute the principal score function by deriving the conditional expectation $\hat h_1(X) :=\hat{E}\{\hat g(X,\hat Z(1))|X\}$ using the distribution $\hat F_{Z(1)|X}$ estimated in Step 1. Generally, numerical integration is required in the calculation of $\hat h_1(X)$. 
\item Estimate $\hat h_1(X_i)$ and $\hat g(X_i,Z_i)$ for each subject in the control group and use Equation (\ref{eq:mu_d_s01_B}) to estimate the treatment effect of interest. For this step, Assumptions A4 and A5 are required.
\end{enumerate}

For the principal stratum $S_{00}$, the prediction for $Y$ and the estimation of the principal scores are combined. For each treatment $T=t$, two consistent estimators for the mean response exist. The estimator provided by Method A is
\begin{equation}  \label{eq:s11A}
\widehat{E}\{Y(t)|S_{00}\} = \frac
{ \sum_{i \in \{i:t_i=1-t, s_i=0\}}  \hat \varphi_t(x_i) }
{ \sum_{i \in \{i:t_i=1-t, s_i=0\}}  \hat h_t(x_i)},
\end{equation}where $\hat \varphi_t (x_i):= \hat{E}[\hat g(X_i,Z_i(t)) \hat \psi_t(X_i,Z_i(t))|X_i=x_i]$ is the expected response for patient $i$ with potential treatment $t$ who can adhere to treatment $t$, conditional on baseline covariate $X_i$. $\hat \varphi_t (X_i)$ can be calculated using three quantities we described how to calculate earlier: $\hat g(X_i,Z_i(t)) $,  $\psi_t(X_i,Z_i(t))$ and $\hat F_{Z(1)|X}$.  

The estimator provided by Method B is
\begin{equation} \label{eq:s11B}
\widehat{E}\{Y(t)|S_{00}\} = \frac
{ \sum_{i \in \{i:t_i=1, s_i=0\}} \hat h_{1-t}(x_i) y_i }
{ \sum_{i \in \{i:t_i=1, s_i=0\}} \hat h_{1-t}(x_i) }.
\end{equation}
The treatment difference for $S_{00}$ can be estimated by any combinations of (\ref{eq:s11A}) and (\ref{eq:s11B}) for the 2 treatments. 

In both Methods A and B, the assumptions A6 and A7 are required in addition to A1-A3 and A4-A5. Although both methods require numerical integration (for marginalizing over conditional expectations of intermediate outcomes in Step 3 for Methods A and B), that can be accomplished using numerical integration packages.  Zhang et al. provide the corresponding unbiased estimation equations for the estimators in (\ref{eq:mu_d_s01_A}), (\ref{eq:mu_d_s01_B}), (\ref{eq:s11A}) and (\ref{eq:s11B}), and the corresponding variance estimation \cite{zhang2020var}. Note that method B can be thought of as a generalization of principal score based methods (Section~\ref{sc.ps.baseline.principal.scores}), therefore  its variance estimator can be applied in this context as well considering a special case of no intermediate variables $Z$.

Importantly, while methods in this section require many assumptions, they allow estimating treatment effect in strata without the need to invoke the assumptions of monotonicity or exclusion restriction. Alternatively, the procedures can be implemented by multiple imputation using a procedure similar but more complex compared to the procedure described in Section~\ref{sc.ps.imputation}. For example, for method A, we can implement Step 3 via sampling from conditional distribution of intermediate covariates $\tilde{Z}(0) \sim  \widehat{F}_{Z(0)|X}$, estimated at Step 1, followed by the imputation of potential outcomes $Y_i(0)$ given patient's covariate profiles $X$ from the experimental treatment arm who adhered to treatment, $i\in \{i:T_i=1,S_i=0\}$ via a regression $\tilde{Y}(0) \sim Y(0)|X, \tilde{Z}(0)$. Averaging over imputed $\tilde{Y}(0)$ essentially results in $\hat \phi_0(X)$ averaged over a subset of patients who adhere to experimental treatment, as in the last term of the right-hand part of expression \ref{eq:mu_d_s01_A}. To ensure ``proper imputation'' \cite{rubin1987}, sampling is from Bayesian posterior distributions rather than from conditional distributions with parameters fixed at estimated values.

Table \ref{table_MI_ext} illustrates the imputation process in the situation when the intercurrent event (treatment discontinuation) can occur at three time points: two intermediate time points (indicated with superscripts $^{(1)}$ or $^{(2)}$ for measured variables) and the final time point (indicated with superscript $^{(3)}$). The variables $X$, $Z^{(1)}$, $Z^{(2)}$, $Y$, $S^{(1)}$, $S^{(2)}$ and $S^{(3)}$, but not the randomized treatment, are used in the imputation process. Note that in general each component of $Z^{(k)}$ can be vector-valued thus conveniently allowing us to incorporate intermediate values of the repeatedly measured primary outcome $Y$, as will be illustrated with our example. Here $S^{(k)}=1$ indicates an intercurrent event at time $K=k$ resulting in missing values for $K \ge k$, shown with dots in the table. The randomized treatment is given in the table for concreteness and is not used as a covariate in the imputation model. The purpose of imputation is to apply the relationships observed in treatment arm $t$ to imputing ``true'' missing outcomes for patients randomized to the same arm, and imputing counterfactual outcomes for patients who were randomized to the parallel arm, $1-t$. The table illustrates the set-up for imputing potential outcomes for a specific treatment $t$, and therefore the imputation process needs to be applied twice with a similar setup for each treatment $t=0,1$.

After the potential outcome $Y_i(t)$ and adherence indicator $S_i(t)$ are imputed, a direct estimate of $Y_i(0), Y_i(1)$ and $Y_i(1)-Y_i(0)$ for any principal stratum based on $S_i(t)$ can be obtained for each imputed sample. Then, the final estimate for the principal stratum can be produced by taking the average across all imputed samples.

\begin{table}[h!tb] \centering
\caption{Illustration of the data setup for imputing potential outcomes including intermediate outcomes and principal strata under treatment $T=t$ for patients assigned to treatment $T=1-t$.}
\begin{tabular}{ccccccccc}
\hline
Subject & Randomized Treatment ($T$) & $X$ & $Z^{(1)}$ & $Z^{(2)}$ & $Y$ & $S^{(1)}$ & $S^{(2)}$ & $S^{(3)}$ \\
\addlinespace[0.2cm]
\hline
001 & $t$ & $\checkmark$ & $\checkmark$ & $\checkmark$ & $\checkmark$& 0 & 0 & 0 \\
002 & $t$ & $\checkmark$ & $\checkmark$ & $\checkmark$ & $\cdot$ & 0 & 0 & 1  \\
003 & $t$ & $\checkmark$ & $\checkmark$ & $\cdot$ & $\cdot$& 0 & 1 & 1 \\
$\cdots$ &&&&&&&&\\
101 & $1-t$ & $\checkmark$ & $\cdot$ & $\cdot$ & $\cdot$ & $\cdot$ & $\cdot$ & $\cdot$ \\
102 & $1-t$ & $\checkmark$ & $\cdot$ & $\cdot$ & $\cdot$& $\cdot$ & $\cdot$ & $\cdot$  \\
103 & $1-t$ & $\checkmark$ & $\cdot$ & $\cdot$ & $\cdot$& $\cdot$ & $\cdot$ & $\cdot$  \\
\hline
\end{tabular}
\\
    {\footnotesize \begin{flushleft} \textbf{Description of variables and symbols}: $X$ is the vector of baseline covariates, $Z^{(1)}$ and $Z^{(2)}$ are the vectors of variables measured at two intermediate time points, $Y$ is the outcome for the response variable at the final time point, and $S^{(1)}$, $S^{(2)}$ and $S^{(3)}$ are indicators of intercurrent events at the intermediate and final time points, respectively. $\checkmark$" indicates non-missing data and ``$\cdot$" indicates missing data. The variable "Randomized Treatment" is not used as a covariate in the multiple imputation procedure.
    \end{flushleft} }
\label{table_MI_ext}
\end{table}

The variance for the treatment difference with imputed data can be estimated either using the Rubin's rule based on between- and within-imputation variances \cite{rubin1976inference, barnard1999miscellanea} or using the bootstrap approach \cite{bartlett2020bootstrap}. As the imputation model and the analysis model are not congenial, the former approach provides a conservative estimate of variance. Therefore, bootstrap is recommended for estimating the variance and constructing the confidence interval.    
More details about the multiple imputation procedure when incorporating postbaseline intermediate outcomes can be found in Reference \cite{luo2020mi}. 

The methods described in this section have been applied to the data example (Section \ref{data.example}) that has been used in previous research \cite{qu2020imp, luo2020mi}. Briefly, the following 7 baseline covariates ($X$) that potentially impact treatment adherence were included: age, gender, HbA1c, LDL-C, triglyceride (TG), fasting serum glucose (FSG), and alanine aminotransferase (ALT). The intermediate outcomes at Week 12 ($Z^{(1)}$, a vector of 6 variables) are HbA1c, LDL-C, TG, FSG, and ALT at Week 12, and injection site reaction adverse events (a binary variable) that occurred between randomization and Week 12. The intermediate outcomes at Week 26 ($Z^{(2)}$, a vector of six variables) are HbA1c, LDL-C, TG, FSG, and ALT at Week 26, and injection site reaction adverse events (a binary variable) that occurred between Weeks 12 and 26. The probability of adherence throughout the entire trial for a given treatment  $t$, ($1-S(t)$) was estimated using a multiplicative probability model:
\[
1 - S(t) = (1-S^{(1)}(t))(1- S^{(2)}(t)) (1- S^{(3)}(t)), t \in \{0,1\},
\]
where $S^{(1)}, S^{(2)}, S^{(3)}$ are the intercurrent event indicators at Weeks 12, 26 and 52, respectively. The probabilities for $S^{(1)}(t)=0$, $S^{(2)}(t)=0$ and $S^{(3)}(t)=0$ are modeled via logistic regression using $X$, $Z^{(1)}$ and $Z^{(2)}$ as covariates. The imputation-based approach for estimating potential outcomes under specific treatment $T=t$ is illustrated in Table \ref{table_MI_ext}. 
The estimates of mean HbA1c for each treatment group as well as the mean difference using both the analytic and multiple imputation approaches are provided in Table \ref{table_ace}. Example of SAS code for implementation of this approach using a simulated data set is given in supplemental materials.

\begin{table}[h!tb]
\centering
\caption{Summary of results of the real data analysis for the estimators of treatment effect in HbA1c for the two populations of adherers using proposed methods (Results are from Table 2 of Reference \cite{qu2020imp} and Table 5 of Reference \cite{luo2020mi}).
}
\label{table_ace}
    \begin{tabular}{ccccc}
    \hline
Method & Population & GL (Estimate $\pm$ SE) & BIL (Estimate $\pm$ SE) & Treatment Difference (95\% CI) \\
\addlinespace[0.2cm]
   \cline{2-5}
   \multirow{2}[1]{*}{Analytic Formulas}
& $S_{*0}$   & 7.59 $\pm$ 0.05 & 7.34 $\pm$ 0.03 & -0.25 (-0.35, -0.15) \\
& $S_{00}$   & 7.55 $\pm$ 0.05 & 7.31 $\pm$ 0.05 & -0.24 (-0.37, -0.10) \\
\cline{2-5}
   \multirow{2}[1]{*}{Multiple Imputation}
& $S_{*0}$  & 7.59 $\pm$ 0.05 & 7.33 $\pm$ 0.04 & -0.26 (-0.35, -0.16) \\
& $S_{00}$    & 7.54 $\pm$ 0.05 & 7.30 $\pm$ 0.04 & -0.25 (-0.34, -0.15) \\
\hline
\end{tabular}
{\footnotesize \begin{flushleft} \textbf{Abbreviations}: BIL, basal insulin peglispro; CI, confidence interval; GL, basal insulin glargine; SE, standard error. \end{flushleft}}
\end{table}

\section{Estimators using Bayesian modeling of principal strata and outcomes}\label{sc.ps.bayesian}

As previously discussed, the major challenge of modeling effects in principal strata is that the observed data are mixtures including latent (unobserved) strata memberships; for example, patients on the experimental treatment with $S=0$ are a mixture of \textit{Immune} and \textit{Benefiters}. Bayesian modeling is often a natural choice for estimating parameters of complex mixtures, as unobserved data and (unobserved) parameters are treated on the same footing as random variables, which provides modeling flexibility. This flexibility allows for a more natural specification of identifiability conditions by rendering often implausible deterministic monotonicity assumptions as stochastic. An early application of Bayesian approaches to Principal stratification in the context of estimating the average treatment effect in compliers (CACE) was presented in the seminal work by Imbens and Rubin \cite{imbens1997bayesian} within a general Bayesian framework of estimating causal parameters in randomized experiments. 
Here we present a simple scenario of Bayesian analysis of principal effects in the setting similar to that  considered in Section~\ref{sc.ps.outcomes.binary} (binary outcome), following the approach of Magnusson et al. (2019) \cite{magnusson2019bayesian} (hereafter, MSRS).

Consider four principal strata $U=\{S_{00},S_{01},S_{10},S_{11}\}$, defined in  (\ref{eq.prstrata}) corresponding to the \textit{Immune}, \textit{Harmed}, \textit{Benefiters}, and \textit{Doomed} strata (as listed in Table~\ref{tab:symptom}) with probabilities, $\pi_{00},\pi_{01},\pi_{10},\pi_{11}$, respectively. For a binary outcome $Y$ we have four strata-specific logits of each potential outcome $Y(1),Y(0)$,
\begin{equation}\label{eq.outcome.logits}
 \theta_u(t)=\textrm{logit}\{\textrm{Pr(}Y(t)=1 \mid U=u)\}, t \in \{0,1\},    
\end{equation}
where $ \textrm{logit}(p)=\textrm{log}(p/(1-p))$. Let vector $\omega$ comprise all parameters $\theta$ and $\pi$, with a posterior 
\begin{equation}\label{eq.bayes.model}
p(\omega \mid Y,S,T) \propto  p(Y \mid S, T, \omega) \cdot p(S \mid T,\omega) \cdot p(\omega).
\end{equation}
Here $p(\omega)$ is the prior and $p(Y \mid S,T,\omega)$ is the likelihood of outcome conditional on (partially observable) strata and treatment, and $p(S \mid T,\omega)$ is the likelihood of strata conditional on treatment. Importantly, monotonicity can be implied by using a strongly informative prior, ensuring probability of \textit{Harmed}, $\textrm{Pr(}\pi_{01}=0)$, is very close to 1 and remains close to 1 after computing posterior. 

As shown in MSRS the strata membership probability is given by a Bernoulli distribution
\[
\textrm{Pr(}S=1 \mid T,\omega)=(1-T) \cdot \textrm{Bern}(\pi_{11} + \pi_{10})+T \cdot \textrm{Bern}(\pi_{11} + \pi_{01}).
\]
In words, the probability of an intercurrent event for patients in the control arm is the sum of the probability of \textit{Benefiters} and \textit{Doomed}; the probability of an intercurrent event for the experimental treatment arm is the sum of probabilities of \textit{Harmed} and \textit{Doomed}.

Recall that each of the four groups of observations formed by combinations of $S$ and $T$ is a mixture of two principal strata:
\begin{enumerate}
\item  $\{i: T_i=1$ and $S_i=1\}$ are subjects from either \textit{Doomed} or  \textit{Harmed} strata
\item  $\{i: T_i=1$ and $S_i=0\}$ are subjects from either \textit{Immune} or  \textit{Benefiters} strata
\item  $\{i: T_i=0$ and $S_i=1\}$ are subjects from either \textit{Doomed} or  \textit{Benefiters} strata
\item  $\{i: T_i=0$ and $S_i=0\}$ are subjects from either \textit{Immune} or  \textit{Harmed} strata
\end{enumerate}

The likelihood of outcome $Y$ for patients in each of these four groups is a mixture of Bernoulli distributions in the corresponding strata with mixing  proportions expressed via strata probabilities. For example, for patients in $\{T=1, S=0\}$, the probability of a binary outcome is a mixture of Bernoulli distributions for the \textit{Immune} and \textit{Benefiters}.
\[
\textrm{Pr(}Y=1 \mid T=1,S=0,\omega)=\frac{\pi_{00}}{\pi_{00}+\pi_{10}} \cdot \textrm{Bern}\{\textrm{expit}(\theta_{00}(1))\} +\frac{\pi_{10}}{\pi_{00}+\pi_{10}} \cdot \textrm{Bern}\{\textrm{expit}(\theta_{10}(1))\},
\]
where $\textrm{expit}(\theta)=(1+\exp(-\theta))^{-1}$, and $\theta_{ij}(t)$ is a shorthand for $\theta_{S_{ij}}(t), i,j,t \in \{0,1\}$.
The full likelihood function is obtained by adding contributions from patients across all observed groups. Under monotonicity the \textit{Harmed} stratum is empty, implying that the first group ($T=1, S=1$) contains only subjects from the \textit{Doomed} stratum and the fourth group ($T=0, S=0$) contains only subjects from the   \textit{Immune} stratum. Imposing additional restrictions can further reduce the number of parameters, making estimation easier. For example, the exclusion restriction implies that $\theta_u(1)=\theta_u(0)$ for $u \in \{S_{00},S_{11}\}$ (since potential outcomes $Y(1)=Y(0)$ for \textit{Doomed} and \textit{Immune}). This assumption may be unrealistic in many settings.

The strata probabilities $\pi_{ij}$ can be parameterized via a logit transformation. Enhancing notation slightly allows reference to strata probabilities via a single index $u$,
\begin{equation}\label{eq.prstrata.softwax}
\pi_{u} = \frac{\exp(\alpha_u)}{\sum_{k \in U}\exp(\alpha_k)}, u \in U,
\end{equation}
where the set $U$ is defined in  (\ref{eq.prstrata}). For identifiability, one set of the parameters $\alpha_u$ should  be set to 0.

Parameter estimation can proceed using a well-known data augmentation (DA) procedure (Tanner and Wong, 1987\cite{tannerwang1987DA}). Data augmentation treats unobserved strata membership $U$ as missing data that are imputed at each iteration from Bernoulli distributions conditional on the current draws from  posterior parameters $\omega$. The latter are subsequently updated, given imputed strata membership $U$. 

Flexible Bayesian modeling of probabilities of strata membership simultaneously with outcomes can be further enhanced by incorporating covariates in estimating strata probabilities $\pi_u(x)$ and outcome parameters $\theta_u(t,x)$ which may help better estimate the effects of interest (in terms of both accuracy and precision). Parameterizing strata membership and outcome probabilities via logistic functions in  (\ref{eq.prstrata.softwax}) and  (\ref{eq.outcome.logits}) allows for a straightforward incorporation of covariates. 
\begin{equation}\label{eq.outcome.logits.x}
\begin{aligned}
&  \theta_u(t,x)=\textrm{logit}\{\textrm{\textrm{Pr(}}Y(t)=1 \mid U=u, X=x)\} \\   
&  \pi_{u}(x) = \frac{\exp(x^{T}\eta_u)}{\sum_{k \in U}\exp(x^{T}\eta_k)},
\end{aligned}  
\end{equation}
where $t \in \{0,1\}; u \in U$, as defined in  (\ref{eq.prstrata}), $X=(1,X_1,..,X_p)$ is a vector of covariates and $\eta_u$ is a strata-specific vector of coefficients including intercept.

Other types of outcomes can be modeled within the same framework. Section~\ref{subsc.bayes.example} presents a straightforward extension to a continuous outcome variable for our example from the diabetes trial while taking into account baseline covariates. 

\subsection{Applying Bayesian mixture modeling to the diabetes study}\label{subsc.bayes.example} 
In this section we apply the approach of MSRS to obtain principal stratification estimates of treatment effect for a continuous outcome (change in HbA1c). The Bayesian modeling is a straightforward extension of the binary model to normally distributed outcomes using a mixture of normal distributions in each stratum. We use the notation of Section~\ref{sc.ps.bayesian}. The four principal strata $U=\{S_{00},S_{01},S_{10},S_{11}\}$ correspond to the \textit{Always-compliers}, \textit{Control-only-compliers}, \textit{Experimental-only-compliers}, and \textit{Never-compliers} of Table~\ref{tab:compl2}. Here we focus on  estimating treatment effect for ``Always-compliers'' (adherers). 

As in Section~\ref{sc.ps.bayesian} we use a general likelihood representation (\ref{eq.bayes.model}) with parameter $\omega$ representing the collection of parameters for each principal stratum, $\omega=\{\beta_U, \delta_U,\sigma^2_U, \eta_U \}$. The first three components govern the normal mixture model for the outcome and the fourth component is the multinomial logit for the probability of strata memberships, as shown in the following.
\begin{equation}\label{eq.mixture.normal.x}
\begin{aligned}
& p(Y \mid T=0,S=0,X=x,\omega)=\frac{\pi_{00}(x)}{\pi_{00}(x)+\pi_{01}(x)} \cdot  \textrm{N}\left( x^T\beta_{00},\sigma^2_{00}\right) +\frac{\pi_{01}(x)}{\pi_{00}(x)+\pi_{01}(x)} \cdot \textrm{N}\left( x^T\beta_{01},\sigma^2_{01} \right) \\
& p(Y \mid T=1,S=0,X=x,\omega)=\frac{\pi_{00}(x)}{\pi_{00}(x)+\pi_{10}(x)} \cdot  \textrm{N}\left( x^T\beta_{00}+\delta_{00},\sigma^2_{00}\right) +\frac{\pi_{10}(x)}{\pi_{00}(x)+\pi_{10}(x)} \cdot \textrm{N}\left( x^T\beta_{10}+\delta_{10},\sigma^2_{10} \right) \\
& p(Y \mid T=0,S=1,X=x,\omega)=\frac{\pi_{11}(x)}{\pi_{11}(x)+\pi_{10}(x)} \cdot  \textrm{N}\left( x^T\beta_{11},\sigma^2_{11}\right) +\frac{\pi_{10}(x)}{\pi_{11}(x)+\pi_{10}(x)} \cdot \textrm{N}\left( x^T\beta_{10},\sigma^2_{10} \right) \\
& p(Y \mid T=1,S=1,X=x,\omega)=\frac{\pi_{11}(x)}{\pi_{11}(x)+\pi_{01}(x)} \cdot  \textrm{N}\left( x^T\beta_{11}+\delta_{11},\sigma^2_{11}\right) +\frac{\pi_{01}(x)}{\pi_{11}(x)+\pi_{01}(x)} \textrm{N}\left( x^T\beta_{01}+\delta_{01},\sigma^2_{01} \right) \cdot
\end{aligned}
\end{equation}
Here $X=(1,X_1,..,X_p)$ is a column vector of unity and $p=7$ covariates: age, gender, HbA1c, LDL-C, triglyceride (TG), fasting serum glucose (FSG), and alanine aminotransferase (ALT), $\beta_{ij} \equiv \beta_{S{ij}}$ are the associated coefficients for the $S_{ij}$ strata, $i,j \in \{0,1\})$.

The mixing proportions $\pi_{ij}(x) \equiv \pi_{S_{ij}}(x)$ are principal stratum probabilities modeled as functions of covariates via the softmax function as expressed in \eqref{eq.outcome.logits.x}. For identifiability, we set all the elements of $\eta_{10}=0$.
The marginal distribution for the probability of the post-randomization event takes the form
\begin{equation}
    \textrm{Pr(}S=1|T=t, X=x,\eta_U)=(1-t) \cdot \textrm{Bern}\left(\pi_{10}(x)+\pi_{01}(x)\right)+t \cdot \textrm{Bern}\left(\pi_{11}(x) +\pi_{01}(x)\right) \nonumber.
\end{equation}

To complete the Bayesian model, we define prior distributions for the parameters. We assume that all the priors are independent of each other.  For analysis of data from diabetes example we assign priors as follows. For each component of coefficient vectors $\beta_U$, $\beta_{ijk} \sim N(0,10), i,j=0,1; k=1,...,(p+1)$. The same prior distribution is also assigned for treatment effects, $\delta_{ij} \sim N(0,10)$. The standard deviations are assigned a $\sigma_{ij}\sim Unif(0.01,20)$ prior. To satisfy the monotonicity assumption we place a strongly informative prior on $\eta_{01}$ to ensure that the probability $\pi_{01}$ (\textit{Control-only compliers}) is very close to 0. Specifically, for the intercept $\eta_{011} \sim N(-50,0.1)$, and for the slopes $\eta_{01k} \sim N(0,1), k=2,..,(p+1)$. Noninformative $N(0,1)$ priors are used for each of the terms in $\eta_{11}$ and $\eta_{00}$.

\begin{figure}[ht]
  \centering
  \includegraphics[scale=0.63, angle=-90]{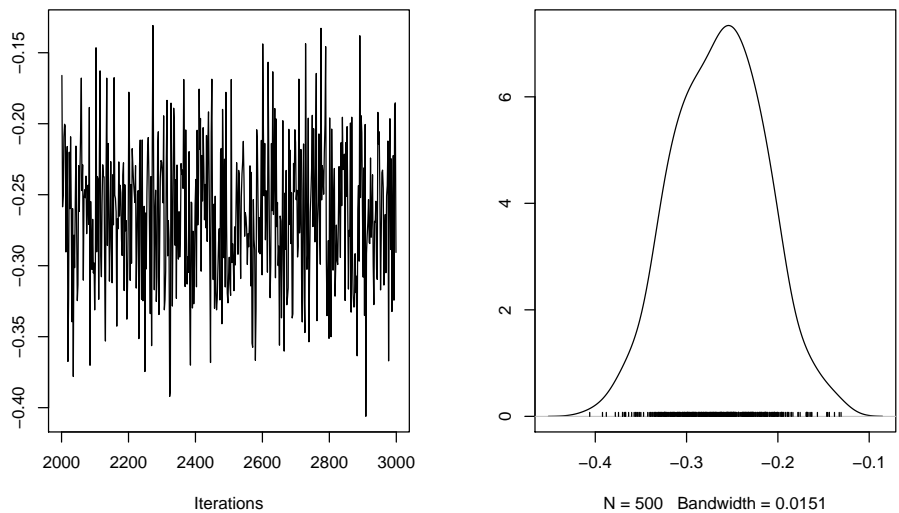}
  \caption{MCMC diagnostics (trace) and posterior density for the treatment effect in \textit{Always-compliers} stratum, $\delta_{00}$.}
  \label{fig.ex1.Bayes}
\end{figure}

Figure~\ref{fig.ex1.Bayes} shows the trace plot from 2000 iterations after 1000 burn-ins (on the left) and posterior density (on the right) for the treatment effect in the ``Always-compliers'' stratum, captured by the parameter $\delta_{00}$. The posterior mean and associated 95\% credible interval are $-0.261$ and $(-0.356, -0.164)$, respectively, which agrees with the results from methods of Section~\ref{sc.ps.baseline.post-baseline}. As a sensitivity analysis to assess potential violations of the monotonicity assumption, we can modify the prior for $\eta_{011}$, as suggested in MSRS. Table~\ref{table.bayes.sens} presents posterior means and 95\% credible intervals for different sets of priors that depart from our initial prior by shifting the mean of the prior distribution toward zero while making it more diffuse. The results for the estimate of $\delta_{00}$ are virtually the same across these scenarios. The last two columns report the posterior mean for the probability of strata membership in ``control-only compliers'', $\pi_{01}$, marginalized over the distribution for covariates, and the associated 95\% credible intervals. Despite diffuse priors centered close to 0 for $\eta_{011}$, the posterior distribution for $\pi_{01}$ remains small. This suggests there is sufficient evidence in the data for a negligible effect of the $01$ strata in estimation of the parameters for other strata of interest.

\begin{table}[h!tb]
\centering
\caption{Sensitivity analysis of monotonicity assumption for estimating $\delta_{00}$ using Bayesian mixtures. Diabetes example.}
\label{table.bayes.sens}
\begin{tabular}{ccccc}
\hline
Prior for $\eta_{011}$ & Posterior Mean for $\delta_{00}$ & 95\% Credible Interval & Posterior mean for $\pi_{01}$ & 95\% Credible Interval\\
\addlinespace[0.2cm]
\hline
N(-50, .1)  & -0.261 & (-0.356, -0.164) & $2.59 \times 10^{-15}$ & $(3.51 \times 10^{-24}, 5.87 \times 10^{-17})$\\
N(-25, .2)  & -0.267 & (-0.362, -0.171) & $2.87 \times 10^{-5}$ & $(2.51 \times 10^{-13}, 2.18 \times 10^{-4})$\\
N(-10, .5)  & -0.262 & (-0.359, -0.162) & $4.03 \times 10^{-4}$ & $(6.04 \times 10^{-7}, 2.91 \times 10^{-3})$\\
N(-5, 1)  & -0.261   & (-0.366, -0.170) & $4.59 \times 10^{-3}$ & $(1.57 \times 10^{-4}, 1.64 \times 10^{-2})$\\
N(-2, 2.5)  & -0.258 & (-0.356, -0.152) & $7.01 \times 10^{-3}$ & $(7.79 \times 10^{-4}, 2.19 \times 10^{-2})$\\
N(0, 10)  & -0.261 & (-0.356, -0.175)  & $7.01 \times 10^{-3}$ & $(3.89 \times 10^{-4}, 2.33 \times 10^{-2})$\\
\hline
\end{tabular}
\end{table}

Note that (\ref{eq.mixture.normal.x}) defines a general mixture model for continuous outcome connecting observed outcomes with parameters governing membership in the four latent strata. When principal stratification is based on an ICE that causes all the subsequent outcomes to be missing, as in our diabetes example, the last two equations are not relevant. Therefore, the regression parameters $\beta_{11},\sigma^2_{11}, \delta_{11}$, and $\pi_{11}$ that solely rely on the outcomes for patients with $S=1$ cannot be estimated. We adopt a modified version of Bayesian principal stratification\cite{magnusson2019bayesian} based on a conditional likelihood model using only subjects with $S=0$. This results in a reduced three-component mixture model where the inestimable parameters are removed from the model. Additional details and R code using package \textbf{R2jags} are provided in supplementary material.

For strata based on ICEs that cause missing outcomes, the general approach of Bayesian mixture modeling\cite{magnusson2019bayesian} may need substantial modification for estimating treatment contrasts in some principal strata. For example, estimating a hypothetical estimand for those who would adhere to the experimental treatment $E(Y(1)-Y(0)|S_{*0})$ would require modeling unobserved  outcomes for patients in the control group who were non-adherent in their group but would have been adherent if randomized to the experimental treatment. This would also require making appropriate assumption(s) about missingness mechanism (e.g MAR) and incorporating earlier observed measures of the outcome $Y$ (or/and possibly additional post-baseline outcomes) for such patients, which is beyond the framework outlined in Section~\ref{sc.ps.bayesian}.

\subsection{Evaluating distributions of baseline covariates within principal strata }\label{sc.cov.in.PS}

Given the latent nature of principal strata, it is informative to evaluate and present  distribution of baseline covariates within PS of interest. For example, Baiocchi et al. (2014)\cite{Baiocchi2014} derived an estimator for the expected value (or proportion) of a single covariate within PS using instrumental variables. Here we provide a more general framework to estimate the within strata distributions for covariates of any type based on estimated probabilities of strata membership, conditional on all available covariates. 

A natural way for obtaining the distribution of covariates within a principal stratum, e.g ``Always compliers'' is using Bayes Rule, $p(X_j | U=S_{00})=w_j p(X_j)$, where $p(X_j )$ is the density of covariate $X_j$ in the overall population and $w_j=\frac{\text{Pr(}U=S_{00} \mid X_j)}{\text{Pr(}U=S_{00})}$  is a covariate-specific weight based on the marginal probability of strata membership given covariate of interest $X_j$. 

To compute weights, we can utilize estimated stratum membership probabilities for each subject $\widehat{\pi}_{00}(x)=\widehat{\text{Pr}}_i(U=S_{00} | X=x)$ using the posterior means from our Bayesian mixture model, where $X=(X_1,..,X_p)$ represents all available covariates. Computing marginal weights requires integrating out all covariates but $X_j$ using multivariate distribution of $X$ which is unwieldy, especially when covariates present a mix of both categorical and continuous variables. In practice, we can use empirical estimates of marginal weights instead of computing it by numerical integration. For a categorial covariate $X_j$, we compute marginal weights associated with each category $l=1,\cdots,L$ as
\[
w_{jl} \approx \frac{\sum_{i=1}^N I(X_{ij}=l)\widehat{\text{Pr}}_i(U=S_{00} \mid X=x_i)/\sum_{i=1}^N I(X_{ij}=l)}{\sum_{i=1}^N\widehat{\text{Pr}}_i(U=S_{00} \mid X=x_i)/N}.
\]
Therefore, we estimate within-strata probabilities,
\[
\widehat{\text{Pr}}(X_j=l \mid U=S_{00}) \approx \frac{\sum_{i=1}^N I(X_{ij}=l)\widehat{\text{Pr}}_i(U=S_{00} \mid X=x_i)}{\sum_{i=1}^N\widehat{\text{Pr}}_i(U=S_{00} \mid X=x_i)}.
\]
It is easy to see that re-weighted proportions $\widehat{\text{Pr}}_i(X_j=l | U=S_{00}),l=1,\ldots,L$  sum to 1 over $L$ categories.

For a continuous covariate $X_j$, each patient is assigned an individual weight, 
\[
w_{ji} \approx \frac{\widehat{\text{Pr}}_i(U=S_{00} \mid X=x_i)}{\sum_{i=1}^N\widehat{\text{Pr}}_i(U=S_{00} \mid X=x_i)}.
\]
We then obtain the density of $X_j$ within principal stratum via weighted kernel density estimation, assuming $\text{Pr}(U=S_{00} | X_j=x_{ij})$ can be approximated by $\text{Pr}(U=S_{00} | X=x_i)$. If the analysis data set contains a large number of subjects having the same value $x_{ij}$ of covariate $X_j$, averaging over individual weights $w_{ji}$  conditional on $X$ would naturally marginalize over all covariates but $X_j$; otherwise we essentially rely on a ``single value'' estimate of the marginal distribution $\text{Pr}(U=S_{00} | X_j)$.

\begin{figure}[ht]
  \centering
  \includegraphics[scale=0.63]{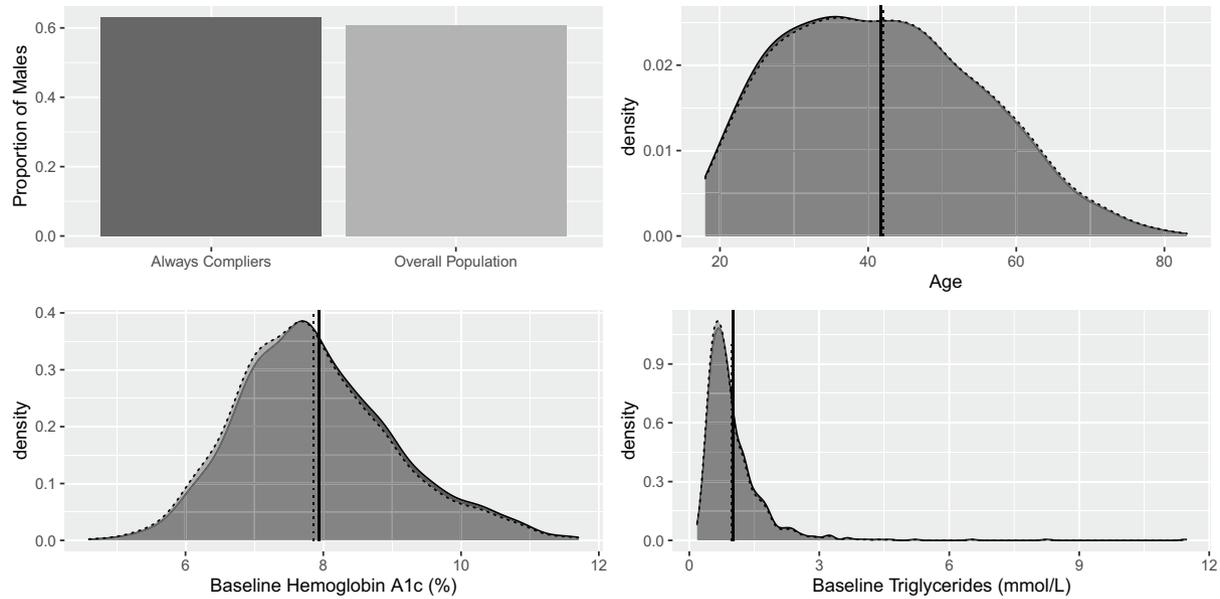}
  \caption{Distribution of age, gender, baseline HbA1c and Triglycerides within ``Always compliers'' stratum (shown in dark gray) and the overall population (light gray). The dotted and solid vertical lines for continuous variables represent the means in the overall population and ``Always compliers'', respectively}
  \label{fig.baseline}
\end{figure}

This approach can accept conditional strata membership probabilities estimated using any method, not necessarily a Bayesian mixture model, as in our case. We illustrate computations of the within-stratum covariate distributions for the diabetes study in Figure~ \ref{fig.baseline}. The graph displays re-weighted gender (proportion of males) and weighted kernel densities for 3 continuous variable at baseline (age, HbA1c and Triglycerides) in the overall population and principal strata based on ``Always compliers'', respectively. The means of continuous variables are represented by vertical dotted (overall population) and solid (``Always compliers'') lines. The ``Always complies" strata is representative of the general population (of this clinical trial) with virtually  identical summaries across the two populations. Similar  pattern was observed for other baseline covariates (not shown).

\section{Extending principal stratification to studies with nonrandomized treatments}\label{sc.ps.nonrand} 
Principal stratification uses the framework of potential outcomes (initially developed in the causal inference literature for analysis of observational data) as a strategy for dealing with intercurrent events in RCTs. The PS strategy has been subsequently adopted in the context of  studies with non-randomized treatments when interest is in causal effects within sub-population(s) defined by early outcomes observed after treatment initiation. In epidemiological studies interest may be in evaluating the causal effect of a non-randomized treatment in a hypothetical subpopulation of patients where a mediator of treatment effect is ``hold'' at the same level. As an example, in a study of smoking ($T$) on lung cancer ($Y$), part of the total effect of smoking may be mediated via high blood pressure $(S=1)$. Then an investigator may be interested in evaluating a direct effect of smoking unmediated by blood pressure. This can be done by evaluating the effects $E(Y(1) - Y(0))$ within principal strata of those who would always have high blood pressure $\{S(0)=1,S(1)=1\}$ and of those who would always have low blood pressure $\{S(0)=0,S(1)=0\}$\cite{chiba2011msm}.  

In nonrandomized trials, the assumptions of SUTVA (implying consistency between observed and potential outcomes) would typically hold, however the treatment ignorability assumptions would not, because treatments are typically assigned by prescribers to optimize expected outcomes. Hence, the expected potential outcome $Y(1)$ is likely to be superior in patients who were assigned to $T=1$ compared to those assigned to $T=0$, i.e $E(Y(1)|T=1) > E(Y(1)|T=0)$. Similarly, $E(Y(0)|T=1) < E(Y(0)|T=0)$. Because treatment assignments are made based on information available to physicians ($X$), we hope that by conditioning on this information (e.g., by stratifying on $X$), the POs become independent of treatment assignment and we can therefore proceed with ``standard analyses'' after conditioning on $X$. This foundational assumption in causal inference is called the ``strong ignorability of treatment assignment'' (Rosenbaum and Rubin\cite{rosenbaum1983propscore}) or ``no unmeasured confounders'' (in the sense that the data available for analysis contain all relevant variables affecting treatment assignment and outcomes). Specifically, we assume 
\begin{equation}\label{eq.NUC}
\begin{aligned}
\mbox{A4*: } \; T \ind \{Y(0),Y(1)\} \mid X. \\
\mbox{A5*: } \; T \ind \{S(0),S(1)\} \mid X.
\end{aligned}
\end{equation}

Several approaches have been used to incorporate covariates in the PS analysis for data with nonrandomized treatments, either directly or by employing propensity-based methods. 

A straightforward extension of sensitivity analyses for PS was undertaken by Egleston et al. (2007) \cite{egleston2007} in the context of estimating SACE. These ideas are illustrated using the example from  Section~\ref{sc.ps.outcomes.binary}. Recall that under randomized treatments and assuming monotonicity we can rewrite the numerator and denominator of the right-hand part of  (\ref{eq.sens.tau}) in terms of observables: 
\begin{equation}
\begin{aligned}
& \nonumber \textrm{Pr(}Y(1)=1 \mid I \; or\; B)=\textrm{Pr(}Y(1)=1 \mid S(1)=0) =\textrm{Pr(}Y=1\mid T=1, S=0) \\ 
& \textrm{Pr(}I \mid I \; or\; B)= \textrm{Pr(}S(0)=0 \mid S(1)=0)=\frac{\textrm{Pr(}S=0 \mid T=0)}{\textrm{Pr(}S=0 \mid T=1)}.  
\end{aligned}
\end{equation}
In nonrandomized trials, assuming ``no unmeasured confounders'' (\ref{eq.NUC}) and using the properties of conditional expectations we can write     
\begin{eqnarray}
\nonumber
\nonumber \textrm{Pr(}Y(1)=1 \mid S(1)=0) &=& 
\nonumber \frac{E_X(\textrm{Pr(}Y(1) \mid S(1)=0, X)\textrm{Pr(}S(1)=0 \mid X))}{E_X(\textrm{Pr(}S(1)=0 \mid X))} \\ &=&
\nonumber \frac{E_X(\textrm{Pr(}Y \mid S=0, T=1, X)\textrm{Pr(}S=0 \mid T=1, X))}{E_X(\textrm{Pr(}S=0 \mid T=1, X))}.
\end{eqnarray}
Similarly, 
\begin{eqnarray}
\nonumber
\nonumber \textrm{Pr(}S(0)=0 \mid S(1)=0) &=& 
\nonumber \frac{\textrm{Pr(}S(0)=0)}{\textrm{Pr(}S(1)=0)} \\ &=&
\nonumber \frac{E_X(\textrm{Pr(}S(0)=0 \mid X))}{E_X(\textrm{Pr(}S(1)=0 \mid X))} \\ &=&
\nonumber \frac{E_X(\textrm{Pr(}S=0 \mid T=0, X))}{E_X(\textrm{Pr(}S=0 \mid T=1, X))}. 
\end{eqnarray}

Letting $g_t(X)=\textrm{Pr(}S=0 \mid T=t, X)$ and $h_t(X)=\textrm{Pr(}Y \mid S=0, T=t, X), t=0,1$, we can write Eq. (\ref{eq.sens.tau}) as 
\begin{equation}\label{eq.sens.tau.obs}
\textrm{Pr(}Y(1)=1 \mid I)=\frac{\frac{E_X(h_1(X)g_1(X))}{E_X(g_1(X))}}{\tau+(1-\tau)\frac{E_X(g_0(X))}{E_X(g_1(X))}},
\end{equation}
where the sensitivity parameter $\tau$ is the risk ratio for \textit{Benefiters} vs. \textit{Immune} strata. Estimation of PS effects in nonrandomized studies critically depends on how well we can estimate $g_t(X)$ and $h_t(X)$ and on the absence of unmeasured confounders\cite{schwartz2012SA}. 

The methods described in Section ~\ref{sc.ps.baseline} can be easily modified for nonrandomized treatments by extending the set of covariates to satisfy assumption  (\ref{eq.NUC}) and integrating out covariates from the marginal probabilities of strata membership. For example, the estimator of $f_1(y|I)$ in the ``strata propensity weighted estimator'' of Section~\ref{sc.ps.baseline.prop.weighted} takes the form 
\[
f_1(y|I)= \frac{E_X(f(y \mid X=x)\textrm{Pr(}S=0 \mid X, T=0))}{E_X(\textrm{Pr(}S=0 \mid T=0,X ))}. 
\]
Estimating this quantity requires modeling both outcome $Y$ and probability  of strata membership as a function of covariates.

Another route is probability of treatment (propensity) weighted estimators. The derivation of such an estimator is outlined below. It may be instructive to compare it with the derivation of the strata propensity weighted estimator in Section~\ref{sc.ps.baseline.prop.weighted}.

\begin{eqnarray}
\nonumber
f_1(y \mid I) &=& f_{Y(1)}(y \mid S(0)=0) \\ &=&  
\nonumber \int_x f_{Y(1)}(y \mid S(0)=0, T=0, X=x)f(x \mid S(0)=0, T=0)dx \\ &=&
\nonumber \int_x f_{Y(1)}(y \mid S(1)=0, T=1, X=x)f(x \mid S=0, T=0)dx \\ &=&
\nonumber \int_x f_{Y}(y \mid S=0, T=1, X=x) \frac{\textrm{Pr(}T=0 \mid S=0, X=x)\textrm{Pr(}S=0 \mid X=x)f(x)}{\textrm{Pr(}S=0 \mid T=0)\textrm{Pr(}T=0)}dx \\ &=&   
\nonumber \frac{E_X(f(y \mid S=0,T=1,X=x)\textrm{Pr(}T=0 \mid S=0, X=x)\textrm{Pr(}S=0 \mid X=x))}{\textrm{Pr(}S=0 \mid T=0)\textrm{Pr(}T=0)}.
\end{eqnarray}

An inverse probability of treatment estimator is constructed in Chiba (2011)\cite{chiba2011msm} using a similar derivation employing the treatment ignorability assumption. However, the principal ignorability assumption is not addressed. This assumption is critical for  transitioning from line 2 to line 3 (allowing us to replace $S(0)$ with $S(1)$ in the conditional statement), whereas treatment ignorability allows us to change conditioning of potential outcomes from $T=0$ to $T=1$ in line 3. We can therefore replace all potential outcomes with observables in line 4 and proceed with various standard statistical modeling. The key element here is introducing inverse probability of treatment weighting given covariates and strata membership, and strata given covariates. 

When the number of potential confounders is large, researchers often use propensity score methodology introduced in Rubin and Rosenbaum (1983) \cite{rosenbaum1983propscore} for obtaining unbiased estimates of treatment effects when analyzing non-randomized studies. The propensity scores are the ``true'' treatment assignment probabilities for each patient. In studies with two treatment arms, as in our case, the propensity score is defined as the conditional probability of receiving treatment $T=1$, given observed baseline confounders $X$: $\pi(X)=\textrm{Pr(}T=1|X=x)$. The fundamental property of propensity is that it is a balancing score in the sense that conditional on $\pi(X)$, the distribution of covariates is  identical (``balanced'') by treatment arms. We can therefore replace the assumption (\ref{eq.NUC}) with  (\ref{eq.NUC.ps}). This essentially means that instead of conditioning on all confounders in the set $X$ we condition on a single-dimensional variate $\pi(X)$, thus replacing a large set of variables with a single variable. 
\begin{equation}\label{eq.NUC.ps}
\begin{aligned}
\mbox{A4*: } \; T \ind \{Y(0),Y(1)\} \mid \pi(X) \\
\mbox{A5*: } \; T \ind \{S(0),S(1)\} \mid \pi(X).
\end{aligned}
\end{equation}
To ensure the validity of propensity score-based methods, the assumptions of SUTVA, strong ignorability (\ref{eq.NUC.ps}) and positivity (that the true propensities $\pi(X)$ are bounded away from 0 and 1 for any  realized configuration $x \in X$) must hold.  

In nonrandomized studies propensity scores are typically unknown and must be estimated from the data (e.g., using logistic regression). There is a rich literature on estimating propensity scores from observational data (see, for example, chapter 4 of Faries et al., 2020\cite{faries2020rw} and references therein). After the propensity scores have been estimated, they can be incorporated within the PS framework in a variety of ways. Two common approaches are stratification by propensity score and using propensity scores as a covariate. The latter is especially convenient for methods that incorporate baseline covariates for modeling principal strata. For example, in Bayesian modeling of Section~\ref{sc.ps.bayesian} we can  incorporate propensity scores within the likelihood function by adding it to the set of covariates for modeling the outcome  and the multinomial logit for the probability of strata membership in (\ref{eq.outcome.logits.x}).       

Methods of Section~\ref{sc.ps.baseline.post-baseline} that incorporate intermediate post-baseline outcomes $(Z)$ in the estimation of treatment effect for principal strata can also be extended to non-randomized studies. Here we provide the main results (using notation of Section~\ref{sc.ps.baseline.post-baseline}), see for details Appendix A of Qu et al., 2020\cite{qu2020general}. With the assumptions A1-A7, the estimators are similar to those described in Section~\ref{sc.ps.baseline.post-baseline} except incorporating the inverse probability weighting based on propensity scores. Again $\pi(x)$ is the probability for a patient taking treatment $T=1$ given the baseline value $X=x$ that has to be estimated from the data (i.e. using the logistic regression). The estimators for the treatment difference for $S_{*0}$ and $S_{00}$ are provided in Table \ref{tab:estimators_nonrandomized}. 

\begin{table}[h!tb] \centering
\caption{Estimators for principal strata-based populations in nonrandomized clinical trials}
\begin{tabular}{ccc}
        \hline
        Population & Method & Estimator for the Treatment Difference\\
 \addlinespace[0.1cm]
        \hline\\
        $S_{*0}$ & A & $\frac{ \sum_j \hat \pi^{-1}(x_j) t_j (1-s_j) y_j}
{
\sum_j \hat \pi^{-1}(x_j) t_j (1-s_j)
}
 - 
\frac{\sum_j \hat \pi^{-1}(x_j) t_j (1-s_j) \hat \phi_0(x_j) }
{
\sum_j \hat \pi^{-1}(x_j) t_j (1-s_j)
}$\\
         & B &
$\frac{ \sum_j \hat \pi^{-1}(x_j) t_j (1-s_j) y_j}
{
\sum_j \hat \pi^{-1}(x_j) t_j (1-s_j)
} 
- 
\frac{
\sum_j \{(1-\hat \pi(x_j))\hat g(x_j,z_j)\}^{-1} \hat h_1(x_j) (1-t_j) (1-s_j) y_j}
{
\sum_j \hat \pi^{-1}(x_j) t_j (1-s_j)
}$
\\
        $S_{00}$ & A &
$\frac
{ \sum_j (1-\hat \pi(x_j))^{-1} \hat \varphi_1(x_j) (1-t_j) (1-s_j) }
{ \sum_j (1-\hat \pi(x_j))^{-1} \hat h_1(x_j) (1-t_j) (1-s_j)}
-
\frac
{ \sum_j \hat \pi^{-1}(x_j) \hat \varphi_0(x_j) t_j (1-s_j) }
{ \sum_j \hat \pi^{-1}(x_j) \hat h_0(x_j) t_j (1-s_j)}$
\\
& B & 
$\frac
{ \sum_j \hat \pi^{-1}(x_j) \hat h_0(x_j) t_j(1-s_j)y_j }
{ \sum_j \hat \pi^{-1}(x_j) \hat h_0(x_j) t_j(1-s_j) }
-
\frac
{ \sum_j (1 - \hat \pi(x_j))^{-1} \hat h_1(x_j) (1-t_j)(1-s_j)y_j }
{ \sum_j (1 - \hat \pi(x_j))^{-1} \hat h_1(x_j) (1-t_j)(1-s_j)}$
\\
\addlinespace[0.1cm]
         \hline 
\addlinespace[0.2cm]
 \end{tabular}
 \\{\footnotesize \textbf{Note}: $\pi(x) := \Pr(T=1|X=x)$. Methods A and B and other notation is as  described in Section~\ref{sc.ps.baseline.post-baseline}, $S=1$ indicates intercurrent event.}
\label{tab:estimators_nonrandomized}
\end{table}

We emphasize that PS analyses with observational data highly depend on assumptions of no unmeasured confounder (\ref{eq.NUC}), therefore sensitivity analysis to assess the consequences of departures from the assumed conditions are needed\cite{schwartz2012SA}. Numerous sensitivity analyses have been developed in causal literature to evaluate the robustness of estimating the average treatment effect in presence of unmeasured confounding (see a review of recent approaches in Zhang et al., 2020\cite{zhang_conf2020}), however extending these methods to principal stratification is not trivial and more research is needed. 

\section{Summary and discussion}\label{sc.summary} 
We considered a wide range of methods for evaluating treatment effects in principal strata defined by post-baseline outcomes or changes in treatment. The key challenge is predicting the unobserved strata membership or evaluating individual probabilities of belonging to a PS. Existing strategies use the following elements as building blocks that are common when dealing with data arising from a mixture with latent (unobserved) categorical variables (here, corresponding to principal strata):
\begin{itemize}
\item	Introducing sensitivity parameters that, when specified, make PS identifiable 
\item	Determining bounds on parameters of interest that are consistent with observed data
\item	Utilizing baseline and post-baseline covariates to help identify latent strata 
\item	Using MI to impute missing PO's for the strata and outcomes
\item	Joint modeling of strata and outcomes using Bayesian mixtures.
\end{itemize}

To help the reader navigate through a vast number of approaches Table~\ref{table_summary} provides a summary of methods considered in this tutorial that includes the estimands of interest, key identifiability assumptions, and references to implementation details in the tutorial and relevant publications.

\begin{table}[h!tb]
\centering
\caption{Summary and references to key approaches for Principal Stratification.
}
\label{table_summary}
    \begin{tabular}{llll}
    \hline
Estimands for & Study type & Key$^*$& Implementation/References\\
Principal stratification & (RCT/OS) & Assumptions &   \\
\addlinespace[0.2cm]
   \cline{1-4}
TE in compliers (CACE) & RCT  & M, ER &  Using IV estimator; Eq. (\ref{eq.IV}) of Section~\ref{estimators.3assum}\cite{air1996identification, littlerubun2000causal}. \\
\addlinespace[0.1cm]

Same as previous & RCT  &  &  Sensitivity analysis with multiple parameters; Section~\ref{sc.ps.monotonicity}, Eq. (\ref{eq.sens.CACE})\cite{lou2019estimation, chiba2011simple}.\\
\addlinespace[0.1cm]

TE (PCE) in strata based & RCT  & SAM &  Modeling joint distribution of partial compliances via copulas;\\
on partial compliance  &   &  &  modeling POs conditional on strata via EM or Bayes; Section~\ref{sc.partial.compliance} \cite{Bartolucci2011, Kim2019, Artman2020}.\\
\addlinespace[0.1cm]

TE in survivors (SACE) & RCT  & M, SD & Sensitivity analysis with sharp  bounds on causal effects; Section~\ref{sc.ps.SACE}\cite{zhang2003estimation}.\\
\addlinespace[0.1cm]

Same as previous & OS  &  TI, M &  Sensitivity analysis with a parameter for the risk ratio in \\
 &   &  & \textit{Experimental-only-survivors} vs \textit{Always survivors}; Section~\ref{sc.ps.nonrand}, Eq. (\ref{eq.sens.tau.obs})\cite{egleston2007}.\\ 
\addlinespace[0.1cm]

TE in patients who & RCT  & M & Sensitivity analysis with a parameter for the risk ratio in \\
would (or would not) &  & & \textit{Benefiters} vs \textit{Immune} strata; Eq (\ref{eq.sens.tau}) of Section~\ref{sc.ps.outcomes.binary}\cite{egleston2010tutorial}.\\
experience an event&  & &   \\
\addlinespace[0.1cm]

Same as previous & RCT & M & Sensitivity analysis with a parameter connecting PO of $Y$ with \\
&  & &  probability of strata membership; Eqs. (\ref{GBH.binary}) and (\ref{eq.senspar1}) of Section~\ref{estimators.2assum}\cite{gbh2003sensitivity}.\\
\addlinespace[0.1cm]

Same as previous & RCT & M, PI & Predicted counterfactual response; Eq. (\ref{eq.pred.count.resp}) of Section~\ref{sc.ps.baseline.cf.response}\cite{bornkamp2019estimating}.\\
\addlinespace[0.1cm]

Same as previous & RCT & SI & PS weighted estimators, probability of strata membership\\
&  & &  modeled via baseline covariates; Eqs. (\ref{eq.base.nomon1}), (\ref{eq.base.nomon2}) of Section~\ref{sc.ps.baseline.prop.weighted}\cite{feller2017principal, qu2021mon}.\\
\addlinespace[0.1cm]

Same as previous & RCT & M, PI & PS weighted estimator, probability of strata  membership\\
&  & & modeled using baseline covariates; \\
&  & & Eq. (\ref{eq.ps.weight}) of Section~\ref{sc.ps.baseline.prop.weighted}, Eq. (\ref{eq.pr.score.est}) of Section~\ref{sc.ps.baseline.principal.scores}\cite{ding2017principal, jo2009use, bornkamp2019estimating}.\\
\addlinespace[0.1cm]

Same as previous & RCT & TI, M & Bayesian mixture analysis, Section~\ref{sc.ps.bayesian}\cite{imbens1997bayesian, magnusson2019bayesian} \\
\addlinespace[0.1cm]

TE in compliers  & RCT  & PI, SI & A general framework for evaluating TE in compliers/adherers \\
(in both or one arm) &   &  &  by jointly modeling POs of the response and PS membership \\
&   &  &    using baseline and postbaseline covariates, Section~\ref{sc.ps.baseline.post-baseline}\cite{qu2020imp, qu2020general}.\\
\addlinespace[0.1cm]

Same as previous  & RCT  & PI, SI & Multiple imputation of POs for strata membership $S$ and outcomes $Y$ \\
 &   &  & using baseline and post-baseline covariates, Sections~\ref{sc.ps.imputation},  \ref{sc.ps.baseline.post-baseline}\cite{luo2020mi}.\\
\addlinespace[0.1cm]

Same as previous  & OS  & TI, PI, SI & A general framework for evaluating TE in non-randomized studies\\
 &   &  & for compliers by jointly modeling outcomes and propensity \\
&   &  &   via baseline covariates, Table~\ref{tab:estimators_nonrandomized}, Section~\ref{sc.ps.nonrand}\cite{qu2020general}. \\

\hline
\end{tabular}
{\footnotesize \begin{flushleft} \textbf{Abbreviations}: ER, exclusion restrictions; M, monotonicity; OS, observational studies; PCE, principal causal effect; PI, Principal ignorability (cross-world independence of PO for strata $S$ and outcome $Y$ given covariates); PO, Potential outcome; PS, principal strata; RCT, randomized clinical trials; SAM, Strong access monotonicity; SD, Stochastic dominance; SI, strata membership ignorability (cross-world independence of strata $S(1)$ and $S(0)$ given covariates); TE, treatment effect; TI, treatment ignorability (given covariates).
\newline
* We omit the assumptions of positivity and SUTVA as they are required for all approaches.
\end{flushleft}}

\end{table}

Given the increased interest in PS following the ICH E9(R1) Addendum, it is important to caution against ``automated'' use of PS estimation ignoring its reliance on untestable assumptions, the need to choose plausible assumptions, and to do sensitivity analyses\cite{Scharfstein2019}. Indeed, methods for dealing with principal strata rely on strong and untestable assumptions. Conceptually, two schools of thought, or camps, exist.

Researchers of the first ``camp'' prefer methods that do not incorporate explicit sensitivity parameters. For example, methods involving baseline or/and post-baseline covariates make ignorability assumptions, implying ``no unmeasured confounders'' (drawing an analogy with the assumption of treatment ignorability in modeling observational data). In the context of principal stratification, this means loosely speaking that all variables predictive of \textit{both} principal strata membership and potential outcomes are included in the analysis. A common drawback of such methods is that sensitivity analyses are often unclear and may not stress-test the main analysis sufficiently. Sometimes researchers propose as ``sensitivity analyses'' variations on the main analysis by incorporating  additional covariates or changing a modeling framework (e.g. from parametric to non-parametric). Such modifications cannot be considered true sensitivity analyses as they do not evaluate consequences of departing from the untestable assumptions. Although sparse, there is literature on sensitivity parameters that directly target departures from assumptions of principal ignorability\cite{ding2017principal, ding2011identifiability}.

Researchers from the second ``camp'' prefer methods incorporating sensitivity parameter(s) naturally leading to sensitivity analyses by varying sensitivity parameters within plausible ranges. This approach avoids relying on covariates that would require untestable ignorability assumptions. As a result, they may fail to utilize potentially relevant information. In principle, any method that we considered can utilize covariates leading to better inference. Avoiding assumptions leads to reliance on multiple sensitivity parameters whose plausibility may be hard to interpret\cite{lou2019estimation}. It may be tempting to present an analysis with sensitivity parameters fixed at specific values (e.g. elicited from experts) not as sensitivity but as the main analysis. This leads to over-confidence in the reported results and, in our opinion, should be avoided. 

Here we advocate for balanced approaches that combine elements of both ``camps'': (1) utilizing all available data via (2) making untestable identifiability assumptions while (3) providing tools for sensitivity analyses testing the robustness of inference to these assumptions.

We note the diversity of frameworks for principal strata may lead to considering an approach accepted in one area as a general standard across areas. For example, it may be tempting to take as a standard setup the PS modeling in compliers as in early work by Imbens and Rubin \cite{imbens1997bayesian} and Robins and colleagues \cite{Robins1998}. This approach is based on a rigid assumption that lack of compliance implies switching to an alternative treatment. This assumption may be reasonable in designs evaluating specific treatment strategies but this assumption is unrealistic for many randomized studies. One could argue that in placebo-controlled studies lack of compliance for patients randomized to active treatment implies switch to placebo. This however ignores the placebo effect and the fact that patients may have access to other treatments. Another questionable assumption (in the context of RCT) is that patients in a control arm may switch to experimental treatment, unless it is permitted by the protocol, e.g., as part of a rescue strategy. However, then it becomes part of a preplanned treatment regimen and is not non-compliance (see Hern\'an and Scharfstein, 2018 \cite{HernanScharfstein2018}). For such designs, PS can be more naturally based not on complier status, but on those who would not require a rescue treatment on both or one of the arms. Targeted approaches for strata based on explicitly defined safety and efficacy criteria have advantage over vague notion of ``compliers'' entertained in the early work on PS. On the other hand, PS defined by treatment completion status $S$ is also useful and here $S=1$ (not completing assigned treatment) simply means that a patient terminated the assigned  treatment regimen at some point. Terminating means either dropping out from the study or switching to any other treatment regimen. Again, ``assigned treatment'' may include dynamic elements such as rescue. A rescued patient may or may not be considered as terminating the initial regimen.

Although potential outcomes are useful for defining and testing causal hypotheses, it is not without controversy. One issue is the deterministic nature of POs in the sense that the outcomes $Y_i(t)$ are ``pre-assigned'' to the patient $i$ under each alternative treatment $t \in \{0,1\}$. Therefore, while $Y(0),Y(1)$ are random variables distributed \textit{across} patients, there is no variability expected \textit{within} for a given patient. Although a philosophical discussion of POs is not within the scope of this tutorial (see, for example Dawid, 2000 \cite{dawid2000}), it is important to note that a stochastic version of POs was introduced in causal literature (see VanderWeele and Robins, 2012 \cite{vanderweele2012stoch} and references therein). 

A stochastic PO framework allows for inherent variability in potential outcomes associated with the same patient to be realized when ``assigning'' an outcome to a patient, similar to the role played by a subject-specific random effect in repeated measures analyses. This is especially relevant for the notion of monotonicity discussed in Section~\ref{sc.ps.monotonicity}. It can be argued that because of its deterministic nature, a commonly accepted monotonicity assumption is implausible although in some settings the probability of violating monotonicity may be small. For example, it may be unlikely to observe weight gain for a patient randomized to placebo if s/he has not experienced it while receiving an active treatment known to cause weight gain. But many AEs are unrelated to treatment and therefore no AE on treatment does not inform about AE on control. 

One can argue that the notion of deterministic POs is well established in causal literature and is well-known through time and practice. In the more traditional setting of estimating the average treatment effect, $E(Y_i(1)-Y_i(0))$, assuming deterministic $Y_i(t)$ has minimal impact on inferences, whereas treating $S_i(t)$ as deterministic in the context of PS-based inference is more problematic. With PS inference, we have to make ``cross-world'' assumptions to connect potential outcomes $S_i(0), S_i(1)$ on the same patient $i$ as if s/he were living in two parallel worlds, one on treatment, and one on control. The deterministic assumption of monotonicity imposes a rigid relationship between the outcomes in the two worlds requiring $S_i(t) \le S_i(1-t)$. Stochastic monotonicity would instead require that the inequality holds in probability, $\text{Pr}(S_i(t)) \le \text{Pr}(S_i(1-t))$, which is more plausible and can be tested from the data.

Another argument for stochastic potential outcomes when modeling PS is based on well-known measurement error models. While the random errors in the response variable $Y$ do not introduce the bias in the regression coefficient, the random errors in the independent variable $X$ do. In the estimation of the treatment effect within a PS, the PS variable is the independent variable.

When estimating effects in principal strata, the monotonicity assumption can be relaxed if we assume $S(t) \ind S(t-1)|X$ in addition to the ignorability assumption $S(t) \ind Y(1-t)|X$ (see Hayden et al., 2005 \cite{hayden2005estimator}). In a recent work by Qu et al. (2021) \cite{qu2021mon}, the plausibility of this assumption versus monotonicity and $S(t) \ind Y(1-t)|X$ was evaluated using cross-over data from a diabetes trial where it was possible to compare a direct evaluation of treatment effects within PS (under the assumption of no carry-over effects) with indirect assessments based on assumptions that do not require monotonicity. As they found, the direct and indirect assessments were in good agreement, whereas monotonicity was for most outcomes inconsistent with observed data from the cross-over trial. However, as such, the monotonicity assumption is unverifiable as it requires access to PO's for alternative treatments measured on the same patients at the same time. The cross-over trials provide merely a crude approximation to such PO's.

Use of cross-world assumptions in PS analyses is a controversial issue. As discussed in Hern\'an and Robins (2020) \cite{hernan2020book}, the principal effects are not cross-world quantities in themselves, i.e., they do not involve potential outcomes $Y(t,s)$ indexed simultaneously by different treatment interventions, e.g., $Y(t,S(t-1))$. Rather, they involve counterfactuals $Y(t,S(t)) \equiv Y(t)$ in a subset of the population. However, some methods for estimation of principal effects make cross-world assumptions about the independence or relationship between the cross-world potential outcomes, e.g., $S(0) \ind S(1)|X$, to deal with non-identifiability. Some researchers in the causal inference literature\cite{hernan2020book} argue against employing cross-world entities because they cannot be identified or verified from data even in principle. A Single World Intervention Graph (SWIG) framework proposed by Richardson and Robins (2013) \cite{richardson2013} was developed to unify a general theory of causal Directed Acyclic Graphs (DAGs)\cite{pearl2009book} and potential outcomes to facilitate inference in settings where the cross-world effects and assumptions are not needed. However, not all treatment effects of interest can be identified without cross-world assumptions. Using the SWIG framework would not allow a method to relax cross-world assumptions in general, but rather would allow a researcher to check whether a treatment effect of interest can be identified without such assumptions given the observed variables and assumed causal structure of a specific problem\cite{breskin2018}.

Many approaches for PS involve estimating various auxiliary (or nuisance) functions of covariates, which is subject to model misspecification. This opens opportunities for using sophisticated machine learning methods capable of incorporating a large number of predictors by employing regularization to reduce variance and overfitting. For example, the predicted values $\hat{m}$ in (\ref{eq.pred.count.resp}), the weight function in (\ref{eq.pred.count.resp}) or principal scores of Section~\ref{sc.ps.baseline.principal.scores} can be based on a simple regression or constructed using methods such as random forest or gradient boosting. However, regularization aims at optimizing prediction of modeled outcomes, rather than optimizing estimation of causal effects. A significant regularization bias in estimation of causal parameters may occur when machine learning estimators of nuisance parameters are simply plugged into the estimating equations of treatment effect and test statistics like (\ref{eq.pred.count.resp}). We note that in the literature on estimating causal parameters from observational data, significant progress was made recently offering solutions to this problem by cleverly combining machine learning models for nuisance parameters resulting in improved convergence rates while ensuring double robustness for the causal parameter(s) of interest (e.g., double or debiased machine learning approach by Chernozhukov et al., 2018 \cite{chernozhukov2018}). In contrast, to our knowledge, there is no literature on optimal use of machine learning for estimating nuisance parameters in the context of PS. As argued in Ding and Lu\cite{ding2017principal}, similarly to general methods on estimating causal effects from observational data, ``it will be interesting to develop doubly robust estimators under the PI assumptions that are consistent when either the principal score or the outcome model is correctly specified''. This may be also a natural place for extending  methodology of double machine learning \cite{chernozhukov2018} to PS.

The interest in PS-based analysis was spurred by a recent ICH E9 (R1) addendum \cite{international2020harmonised} where PS is presented as one of five strategies for dealing with intercurrent events. As was pointed out in several recent reviews of the guidance\cite{Scharfstein2019, qu2021covid}, PS should be understood as a strategy for \textit{defining a population} of interest (often based on an ICE) rather than as a strategy for \textit{dealing with ICEs}. Clearly, even if we knew exactly which patients belonged to the principal stratum of interest, we would have still needed to define a strategy for dealing with other ICEs and describe potential outcomes for imputing missing data. In many PS methods, it is tacitly assumed that potential outcomes $Y(0), Y(1)$ for patients within the stratum of interest are well defined. One reason why this may be the case is that historically PS strategies were entertained for CACE and SACE estimands where outcomes for those who belong in the stratum of interest are non-missing. However, for outcomes from longitudinal trials, estimating PS effects within arbitrary strata defined by potential outcomes $S(0), S(1)$ requires jointly modeling intercurrent events and outcomes as repeated measures. Section~\ref{sc.ps.baseline.post-baseline} reviews several recently proposed methods that may help bridging this gap in literature.  

This review is limited to PS strategies for continuous or binary outcomes. Estimating treatment effects for time to event outcomes within principal strata can arise in oncology and other areas. This is a challenging task because it requires modeling \textit{competing events}, time to ICE and time to event of interest (see discussion of methods and some examples in Bornkamp et al., 2021\cite{bornkamp2021} and references therein).

Another topic not covered by this tutorial is the use of principal stratification in mediation analysis that generated a large number of publications starting from Rubin (2004)\cite{Rubin2004} attempting to formulate direct and indirect causal effects via PS. Establishing connection between mediation and PS was recognized as a challenging problem in the causal community\cite{vanderweele2011principal, Mealli2012} and different approaches were considered (see a recent research\cite{Kim2019} proposing a unifying Bayesian framework for PS and mediation analyses and references therein).

Although the focus of this tutorial is on RCTs, we covered to some extend applications of PS methods for observational studies. PS methods were developed within a general causal framework of potential outcomes. Therefore, it is not surprising that POs can often be accommodated within the same framework when initial treatment assignment is nonrandom. This requires the additional assumption of treatment ignorability conditional on covariates. Methods that utilize pre-treatment covariates can be used for identifying effects within principal strata while removing bias due to non-random treatment assignment.
Finally, we note a potentially promising approach inspired by so-called \textit{separable treatment effects} introduced in Stensrud et al. (2020) \cite{stensrud2020sepeff}. The same authors \cite{stensrud2020condsepeff} proposed an alternative formulation of causal effect in a stratum defined by post-baseline variables using the concept of conditional separable effects which may attract attention of practitioners in the future. 

\section*{Acknowledgements}
We thank Dr. Stephen J. Ruberg and the DIA Scientific working group on estimands and missing data for numerous discussions, and Dr. Yu Du for reviewing the manuscript and providing valuable comments.



\bibliography{references}

\end{document}